\documentclass[preprintnumbers,amsmath,amssymb,floatfix,10pt,prd,onecolumn,
superscriptaddress,nofootinbib]{revtex4-2}
\usepackage{latexsym}
\usepackage{epsfig}
\usepackage{epstopdf}
\usepackage{graphicx}
\usepackage{amssymb}
\usepackage{amsmath}
\usepackage{amsfonts}
\usepackage{subfigure}
\usepackage{dcolumn}
\usepackage{bm}
\usepackage[normalem]{ulem}
\usepackage[dvipsnames]{xcolor}
\usepackage{hyperref}
\usepackage{color}
\usepackage{comment}
\usepackage{float}
\usepackage[utf8]{inputenc}
\hypersetup{colorlinks=true,linkcolor=blue,citecolor=red,urlcolor=magenta}
\usepackage{mathrsfs}
\usepackage{mathtools}
\usepackage{orcidlink}
\usepackage{appendix}

\begin{document}

\title{{Finite-range hair from a nonlocal matter sector: charged Yukawa black holes}}
\author{Francisco Tello-Ortiz \orcidlink{0000-0002-7104-5746}}
\email{francisco.tello@ufrontera.cl}
\affiliation{Departamento de Ciencias Físicas, Universidad de La Frontera, Casilla 54-D, 4811186 Temuco, Chile.}

\author{Hishan Farfán-Bachiloglu \orcidlink{0000-0002-1382-5605}}
 \email{hishan.farfan@ucn.cl}
\affiliation{Departamento de F\'isica, Universidad Cat\'olica del Norte, Av. Angamos 0610, Antofagasta, Chile.}

\author{Alejandro Rueda
\orcidlink{0009-0005-6302-0038}}
\email{uqarueda@uq.edu.au}
\affiliation{School of Mathematics and Physics, University of Queensland, Brisbane, Queensland 4072, Australia}

\author{M. Zubair
\orcidlink{0000-0003-2227-788X}}
\email{drmzubair@cuilahore.edu.pk}
\affiliation{Department of Mathematics, COMSATS University Islamabad, Lahore Campus, Lahore, Pakistan}

\begin{abstract}
{
We construct a exact charged, static and spherically symmetric black hole solution of the Einstein--Maxwell equations with an additional anisotropic matter sector, whose metric is the Reissner--Nordstr\"om geometry endowed with a Yukawa-type contribution. The Yukawa term is exponentially screened: it is visible in the strong-field region but carries no independent asymptotic charge. We obtain an analytic parametrization of the extremal curve, identify the condition for higher-order degenerate horizons, characterize the curvature singularity and establish the weak, null, dominant and strong energy conditions. We further study the horizon thermodynamics, including the conjugate variables associated with the amplitude and range of the Yukawa sector and the corresponding generalized Smarr relation. The photon sphere, shadow, ray-traced appearance and quasinormal ringdown are then analyzed, with quasinormal frequencies computed through the sixth-order WKB expansion and confirmed, overtones included, by an independent spectral method. {Confronting the model shadow with Event Horizon Telescope measurements of M87* and Sgr A* constrains the finite-range sector directly, favoring its strongly screened, near-Reissner--Nordstr\"om regime.} This provides a unified geometric and phenomenological characterization of finite-range charged black hole hair in ordinary Einstein gravity.
}
\end{abstract}

\maketitle

\section{Introduction}

{
General Relativity (GR) offers a remarkably rigid description of the end state of gravitational collapse. In electrovacuum, stationary black holes belong to the Kerr--Newman family, completely specified by their mass, angular momentum and electric charge \cite{Israel:1967wq,Carter:1971zc,Robinson:1975bv,Kerr:1963ud,Newman:1965my,Chrusciel:2012jk}. This uniqueness underlies the no-hair conjecture \cite{Ruffini:1971bza,Hawking:1971vc}, which has been supported by a series of no-scalar-hair theorems \cite{Bekenstein:1995un,Mayo:1996mv}. The same rigidity is what makes black holes such sensitive laboratories: any additional ingredient that survives in the exterior region, be it a new field, a surrounding medium or a change in the gravitational dynamics, leaves an imprint that can in principle be read from strong-field observations \cite{Berti:2015itd,Cardoso:2016ryw,Barack:2018yly,Bambi:2019xzp,Vagnozzi:2022moj}.

Hair is nevertheless known to arise in several controlled ways. Black holes immersed in matter, the so-called dirty black holes, retain the basic structure of their horizons while their exterior is reshaped by the surrounding distribution \cite{Visser:1992qh}. Scalar fields with suitable couplings or potentials support hairy solutions that evade the classical theorems \cite{Martinez:2002ru,Martinez:2004nb,Herdeiro:2015waa}, and soft hair has been associated even with the horizon of the Schwarzschild geometry \cite{Hawking:2016msc}. On the astrophysical side, black holes surrounded by dark-matter halos or quintessence-like fluids provide analytic examples in which a matter sector modifies the strong-field metric while the asymptotic charges remain those of the vacuum solution \cite{Kiselev:2002dx,Konoplya:2019sns,Konoplya:2022hbl,Heydarzade:2023dof}. More recently, constructions based on effective anisotropic sources have produced hairy and regular black holes whose exterior is controlled by the matter content \cite{Ovalle:2020kpd,Ovalle:2023ref,Avalos:2023ywb}.

Two broad routes lead beyond the standard charged black hole. The first modifies the gravitational dynamics, through $f(R)$, $f(G)$ and $f(T)$ theories \cite{Moon:2011hq,delaCruz-Dombriz:2009pzc,deSousaSilva:2018kkt,Junior:2015fya}, scalar--tensor and vector--tensor models \cite{Kim:1998hc,Gao:2017vqv}, higher-curvature gravity \cite{Konoplya:2008ix,Konoplya:2020bxa,Konoplya:2020juj}, or quantum-gravity-motivated scenarios such as asymptotic safety, scale dependence and quantum-corrected geometries \cite{Bonanno:2000ep,Cai:2010zh,Falls:2010he,Koch:2016uso,Rincon:2018lyd,Casadio:2022ndh}. The second route retains Einstein gravity and enlarges the matter sector, for instance with nonlinear electrodynamics \cite{Nojiri:2017kex,Bargueno:2020ais,Tzikas:2018cvs} or anisotropic fluids. Its advantage is conceptual: it separates the effect of a non-vacuum source from genuinely gravitational modifications, and it makes it possible to ask directly which matter content is compatible with a given strong-field geometry \cite{Carballo-Rubio:2018pmi}.

This question is more demanding than it may appear. In a large fraction of the literature the metric is prescribed first, frequently by inserting an exponential, Yukawa-type or power-law correction into a known solution, and the source is reconstructed afterwards. The matter content then has to be examined a posteriori: whether the energy density has a definite sign, whether the standard energy conditions hold \cite{Hawking:1973uf,Visser:1997tq,Curiel:2014zba}, and how these requirements constrain the number and nature of the horizons \cite{Yang:2021civ}. Yukawa-type black holes have indeed been obtained along several distinct routes. In higher-derivative gravity the massive spin-two mode produces Yukawa terms in the static field \cite{Stelle:1977ry,Stelle:1977gr}, and the corresponding black hole solutions have been studied in detail \cite{Lu:2015cqa,Lu:2015psa}; exponentially suppressed, Yukawa-like corrections also arise in nonlocal Deser--Woodard gravity \cite{Li:2026drw}. A third route starts from a Yukawa-modified gravitational potential, converts it into an effective energy density, and integrates the Einstein equations, with or without electric charge \cite{Gonzalez:2023rsd,AraujoFilho:2023ibe,AraujoFilho:2024ctw}. In each case the Yukawa profile is inherited from an assumed potential or from the extra gravitational degrees of freedom of the underlying theory, rather than from a property of the matter that sources the geometry. The issue has physical consequences well beyond mathematical consistency: the interior structure of charged black holes, the fate of the Cauchy horizon and the mass-inflation instability all depend sensitively on the matter content \cite{Poisson:1989zz,Poisson:1990eh,German:2001ti,Dafermos:2003vim,Hartnoll:2020rwq,Cai:2020wrp,An:2021plu,Devecioglu:2021xug,Ovalle:2023vvu}. A framework in which the profile of the additional sector follows from a constitutive principle within general relativity, rather than from a choice of the metric or of the potential, is therefore highly desirable. Finite-range, Yukawa-type behavior is a particularly well-motivated target: it is ubiquitous in massive and effective descriptions of gravity and matter, it preserves the asymptotic charges, and it provides the simplest setting in which hair is visible in the strong-field region while screened at large distances.

Anisotropic matter offers the natural language for such a construction. Local anisotropy allows the radial and tangential stresses to respond differently to the geometry, and nonlocal equations of state go one step further by letting the pressure at a given radius depend on the matter enclosed within it. Nonlocal closures were introduced in the study of relativistic fluid spheres \cite{Hernandez:1998hk,Hernandez:2001nr} and relate the radial pressure to integrals of the density over the enclosed region, thereby encoding a collective property of the distribution that a purely local equation of state cannot capture. For static spherical symmetry, the Kerr--Schild condition \cite{Kerr:1965wfc,Jacobson:2007tj} ties the radial pressure of the additional sector to its density and therefore constrains the admissible matter content very strongly. Putting these two ingredients together turns the question above into a well-posed one: which radial profile is compatible with both a nonlocal closure and the Kerr--Schild structure, and what geometry does it generate?

In this work we show that this program can be carried out analytically. A finite-range generalization of the nonlocal equation of state, combined with the Kerr--Schild condition, determines the density profile of the additional sector and yields an exact charged black hole of Einstein--Maxwell gravity with a Yukawa-type contribution. The mass and the electric charge remain the only asymptotic charges: for a finite range the new term is hair that is visible in the strong-field region and exponentially screened at large radius, and in the opposite long-range limit it merges continuously with a shift of the mass. The Yukawa behavior is therefore a prediction of the construction and not an input. The geometry is an exact solution of the Einstein--Maxwell equations with an effective anisotropic fluid as additional source, for which the nonlocal closure acts as a constitutive relation that selects the profile.
A solution of this kind is useful only if its consequences can be followed through to observables. We therefore characterize the geometry in a unified way. Its causal structure admits an analytic description of the extremal boundary and of higher-order degenerate horizons even though the horizon equation is transcendental, and the complete set of energy conditions, rather than the sign of the density alone, delimits the branch of parameters with a physically sound source. The horizon thermodynamics, formulated with the amplitude and the range of the Yukawa sector as additional variables, extends the classical laws of black hole mechanics \cite{Bekenstein:1973ur,Bardeen:1973gs,Hawking:1976de,Smarr:1973zz,Iyer:1994ys} and leads to a generalized Smarr relation, in line with recent extended formulations in which couplings are varied \cite{Xiao:2023lap,Hajian:2023cwe}.

The strong-field phenomenology then connects the construction with present and foreseeable data. Gravitational-wave observations probe the relaxation of compact objects through their ringdown spectrum \cite{LIGOScientific:2016aoc,LIGOScientific:2016sjg,LIGOScientific:2017bnn,LIGOScientific:2017vox,LIGOScientific:2017vwq,LIGOScientific:2017ycc}, whose theory is reviewed in Refs.~\cite{Kokkotas:1999bd,Berti:2009kk,Konoplya:2011qq} and which has become a standard diagnostic of deviations from GR \cite{Konoplya:2016jvv}. Horizon-scale imaging by the Event Horizon Telescope, in turn, constrains the optical structure associated with unstable photon motion, namely the photon sphere, the shadow and the photon rings \cite{EventHorizonTelescope:2019dse,EventHorizonTelescope:2019uob,EventHorizonTelescope:2019jan,EventHorizonTelescope:2019ths,EventHorizonTelescope:2019pgp,EventHorizonTelescope:2019ggy,EventHorizonTelescope:2022wkp,EventHorizonTelescope:2022xqj}. Shadows and their image morphology have been studied extensively for modified black holes and wormholes \cite{Luminet1979,Gralla2019,GLM2020,Vincent2022,Perlick:2021aok,Atamurotov:2013sca,Papnoi:2014aaa,Atamurotov:2015nra,Heydari-Fard:2021pjc,Heydari-Fard:2021qdc,Ovgun:2023ego,Macedo:2025ipc}, and quasinormal spectra have been computed for a wide range of regular, hairy and modified geometries \cite{Panotopoulos:2017hns,Panotopoulos:2019qjk,Destounis:2018utr,Gonzalez:2018xrq,Rincon:2018sgd,Li:2022kch,Avalos:2023jeh,Konoplya:2016hmd,Konoplya:2019goy,Konoplya:2023ahd}. Since photon spheres, shadows and quasinormal modes probe the same strong-field region \cite{Chandrasekhar1992,Cardoso:2008bp,Konoplya:2017wot}, their joint analysis is a sharp test of how much of a finite-range sector actually reaches the scattering region. We therefore compute the photon sphere and shadow and confront the latter with the Event Horizon Telescope measurements of M87* and Sgr A*, illustrate the image morphology by ray tracing, and obtain the scalar and electromagnetic quasinormal spectra with a sixth-order WKB expansion validated by an independent spectral method, including the overtones. The eikonal limit then links the optical and the dynamical observables through the angular frequency and the Lyapunov exponent of the unstable null orbit. In this way a single finite-range matter sector is followed from its source, through its causal structure and thermodynamics, to its optical and spectral signatures, within one self-consistent framework.

The article is organized as follows. Sect.~\ref{sec:yukawa} derives the solution and studies its screened mass profile, horizon structure, curvature and energy conditions. Sect.~\ref{sec:thermo} develops the horizon thermodynamics and generalized Smarr relation. Sect.~\ref{sec:optics} addresses the photon sphere, shadow and ray-traced images, including a direct comparison with Event Horizon Telescope shadow measurements. Sect.~\ref{sec:qnm} discusses scalar and electromagnetic test-field perturbations and their eikonal connection with null geodesics. Finally, Sect.~\ref{section8} summarizes the main results and their physical implications.

Throughout the article we use the mostly positive signature $\{-;+;+;+\}$ and geometrized units $c=G=1$, so that $\kappa=8\pi$.
}

\section{Charged black holes with a finite-range matter sector}
\label{sec:yukawa}

\subsection{Nonlocal matter sector and exact screened geometry}

{
We consider a static and spherically symmetric geometry in Schwarzschild gauge,
\begin{equation}
ds^{2}=-f(r)\,dt^{2}+\frac{dr^{2}}{f(r)}+r^{2}d\Omega^{2},
\label{ansatz}
\end{equation}
which satisfies $g_{tt}g_{rr}=-1$, and write $f(r)=1-2m(r)/r$. The matter content is the electric field of a point charge, $E(r)=Q/r^{2}$, together with an additional anisotropic sector of density $\rho_h$ and radial pressure $p_{r,h}$, the subscript $h$ labeling this finite-range (hair) sector throughout. For the Maxwell field $\rho_{\rm EM}=-p_{r,{\rm EM}}=Q^{2}/(8\pi r^{4})$. The $tt$ and $rr$ components of the Einstein equations then give
\begin{align}
8\pi\left[\rho_{\rm EM}(r)+\rho_h(r)\right]&=\frac{2m'(r)}{r^{2}}, \label{rhogen}\\
p_{r,h}(r)&=-\rho_h(r). \label{prh}
\end{align}
Equation~\eqref{prh} is the geometrical consequence of the one-function ansatz, which forces $T^t{}_t=T^r{}_r$ \cite{Jacobson:2007tj}. It is satisfied identically by the Maxwell field and therefore restricts the additional sector alone. Matter-supported black holes of this type belong to the class of ``dirty'' black holes, in which non-vacuum fields modify the exterior geometry while Einstein gravity is left unchanged \cite{Visser:1992qh}. The remaining freedom is encoded in a single radial density profile, which we determine by imposing a nonlocal equation of state on the additional sector.}{
Nonlocal equations of state were introduced in relativistic fluid models to encode the fact that the stress at a given radius can depend on the matter distribution throughout the enclosed region rather than only on local thermodynamic variables \cite{Hernandez:1998hk,Hernandez:2001nr}. This is precisely the structural ingredient needed here. Building on the integral structure introduced in Ref. \cite{Hernandez:2001nr}, we introduce the finite-range generalization
{\begin{equation}
p_{r,h}(r)=\rho_h(r)
+\frac{A}{r^2}\int_0^r x^2\rho_h(x)\,dx
+\frac{\sigma}{r^2},
\label{nleos}
\end{equation}
where $A>0$ is an inverse length and $\sigma$ is a dimensionless constant. Combining Eqs.~\eqref{prh} and \eqref{nleos} gives
\begin{equation}
-2\rho_h(r)=
\frac{A}{r^2}\int_0^r x^2\rho_h(x)\,dx
+\frac{\sigma}{r^2}.
\label{intcon}
\end{equation}
Differentiating Eq.~\eqref{intcon} yields
\begin{equation}
2r\rho_h'(r)+(4+Ar)\rho_h(r)=0,
\label{oderho}
\end{equation}
whose solution is
\begin{equation}
\rho_h(r)=-\frac{Al}{16\pi r^2}\,e^{-Ar/2},
\label{rhoh}
\end{equation}
with $l$ an integration constant of dimension length. Inserting this profile into Eq.~\eqref{intcon} gives
\begin{equation}
\frac{A}{r^2}\int_0^r x^2\rho_h(x)\,dx
=-\frac{Al}{8\pi r^2}\left(1-e^{-Ar/2}\right),
\end{equation}
and the equation is satisfied identically for
\begin{equation}
\sigma=\frac{Al}{8\pi}.
\label{sigmaA}
\end{equation}
The additional sector is therefore completely specified by the pair $(A,l)$.
Using Eq.~\eqref{rhogen}, the mass function of the additional sector obeys
\begin{equation}
m_h'(r)=4\pi r^2\rho_h(r)=-\frac{Al}{4}\,e^{-Ar/2},
\end{equation}
so that $m_h(r)=\tfrac{l}{2}e^{-Ar/2}+m_{h,0}$. For $A>0$ the finite-range sector carries no residual mass at spatial infinity, which sets $m_{h,0}=0$ and gives
\begin{equation}
m_h(r)=\frac{l}{2}e^{-Ar/2},
\qquad
\rho_h(r)=-\frac{Al}{16\pi r^2}e^{-Ar/2}.
\label{mhrhoh}
\end{equation}
The characteristic screening length is $2/A$ in the exponential convention adopted here.
The construction is more restrictive than choosing a Yukawa-type metric and reconstructing a source a posteriori: once the one-function geometry and the nonlocal closure are imposed, the exponential profile is a consequence and not an input. The amplitude $l$ parametrizes the resulting finite-range sector but is not an independent asymptotic charge.}
}
{
Adding the Maxwell contribution,
\begin{equation}
m_{\rm EM}(r)=M-\frac{Q^2}{2r},
\end{equation}
gives the total Misner--Sharp mass
\begin{equation}
m(r)=M-\frac{Q^2}{2r}+\frac{l}{2}e^{-Ar/2},
\label{MSmass}
\end{equation}
with
\begin{equation}
f(r)=1-\frac{2M}{r}+\frac{Q^2}{r^2}
-\frac{l}{r}e^{-Ar/2}.
\label{fmetric}
\end{equation}

For finite $A>0$ the exponential term falls faster than any inverse power. Consequently,
\begin{equation}
m(r)=M-\frac{Q^2}{2r}+{\cal O}\!\left(e^{-Ar/2}\right),
\qquad r\rightarrow\infty,
\end{equation}
and the ADM mass is exactly $M$. The parameter $l$ therefore does not define an additional asymptotic Gauss-law charge. Instead, it changes the quasi-local mass in the strong-field region while becoming exponentially invisible at large radius. In this precise sense the solution carries screened, finite-range hair.

The limiting cases sharpen this interpretation:
\begin{align}
l\rightarrow0 &: \qquad f(r)\rightarrow f_{\rm RN}(r),\\
A\rightarrow+\infty &: \qquad f(r)\rightarrow f_{\rm RN}(r)
\quad (r>0),\\
A\rightarrow0 &: \qquad
f(r)\rightarrow
1-\frac{2(M+l/2)}{r}+\frac{Q^2}{r^2}.
\end{align}
Thus the long-range limit is qualitatively different from every finite-$A$ configuration: the Yukawa amplitude becomes indistinguishable from a mass renormalization,
\begin{equation}
\widetilde M=M+\frac{l}{2}.
\end{equation}
The pair $(A,l)$ therefore controls not only the magnitude of the Yukawa term but also whether it is distinguishable from an asymptotic conserved charge.

This distinction also fixes the sense in which the word ``hair'' is used here. Standard no-hair statements constrain additional long-range field charges under specific assumptions on the matter content and couplings \cite{Bekenstein:1995un,Mayo:1996mv}. Our parameter $l$ is of a different type: for every finite $A>0$ it leaves no independent $1/r$ coefficient at infinity and hence no additional Gauss-law charge. It is instead quasi-local, matter-supported hair whose influence is restricted to radii $r\lesssim 2/A$. The solution is therefore not presented as a counterexample to classical no-hair theorems; rather, it provides an explicit example of how non-vacuum structure can remain dynamically relevant in the strong-field region while being exponentially screened from the asymptotic charge data.
}

The exact solution is now completely specified. The next question is whether
the finite-range contribution merely perturbs RN quantitatively or can also
change its causal organization. Because the horizon equation contains the
exponential and is therefore transcendental, the useful object is not a
closed expression for every root but the analytic locus on which roots merge.

\subsection{Horizon structure and analytic extremal locus}

{
Horizons are positive roots of
\begin{equation}
f(r_H)=0.
\label{horEq}
\end{equation}
The transcendental dependence on $r_H$ prevents a generic closed-form expression for the individual roots. The \emph{extremal boundary}, however, can be characterized analytically. Let $r_\star$ denote a double positive root. The conditions
\begin{equation}
f(r_\star)=0,
\qquad
f'(r_\star)=0
\label{extconditions}
\end{equation}
can be solved algebraically for $(A,l)$ in terms of $r_\star$. Defining
\begin{equation}
\Delta_\star=r_\star^2-2Mr_\star+Q^2,
\end{equation}
one obtains
\begin{align}
A_{\rm ext}(r_\star)
&=
\frac{2(Q^2-r_\star^2)}
{r_\star\Delta_\star},
\label{Aext}\\
l_{\rm ext}(r_\star)
&=
\frac{\Delta_\star}{r_\star}
\exp\!\left[\frac{A_{\rm ext}(r_\star)r_\star}{2}\right].
\label{lext}
\end{align}
Only portions of this parametric curve satisfying $A_{\rm ext}>0$ belong to the finite-range asymptotically flat family considered here. Equations~\eqref{Aext}--\eqref{lext} are more informative than a purely numerical extremality scan: they identify directly how the critical Yukawa amplitude and screening length are correlated with the degenerate horizon radius.

Figure~\ref{newfig1} illustrates the corresponding causal sectors for the representative choice $M=1$ and $Q=0.5$. The extremal curve separates configurations possessing two simple positive roots from horizonless configurations. Importantly, no universal statement based only on the numerical value of $l$ is possible: the number and degeneracy of horizons depend jointly on $(M,Q,A,l)$.

\begin{figure*}
\begin{center}
\subfigure{\includegraphics[width=0.4\textwidth]{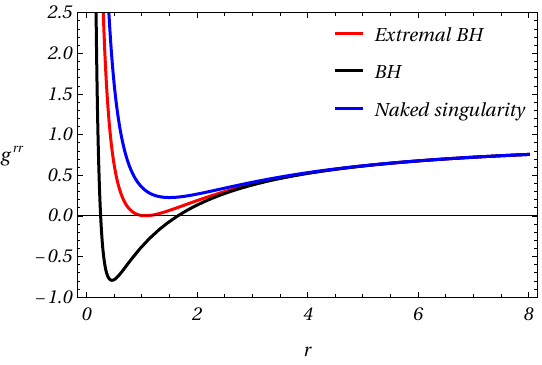}}~~~
\subfigure{\includegraphics[width=0.4\textwidth]{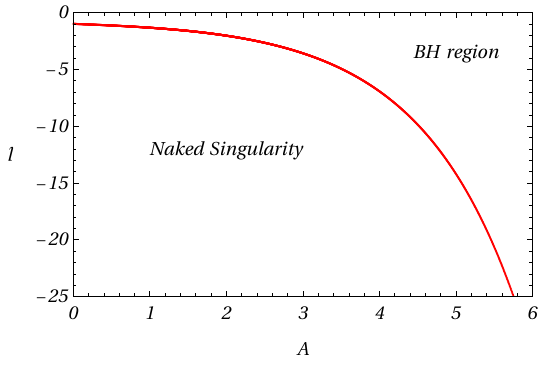}}
\end{center}
\caption{\textbf{Left panel}: Inverse radial metric potential for representative two-horizon, extremal and horizonless configurations. \textbf{Right panel}: Black hole and horizonless sectors in the $(l,A)$ plane for $M=1$ and $Q=0.5$. The critical boundary can equivalently be generated from the analytic parametrization \eqref{Aext}--\eqref{lext}.}
\label{newfig1}
\end{figure*}

The analytic parametrization also allows one to search for a higher-order degeneracy. Imposing in addition
\begin{equation}
f''(r_\star)=0
\end{equation}
and using Eqs.~\eqref{Aext}--\eqref{lext} reduces the triple-root condition to the algebraic equation
\begin{equation}
r_\star^4-4Q^2r_\star^2+4MQ^2r_\star-Q^4=0.
\label{tripleroot}
\end{equation}
Therefore, higher-order degeneracy is not excluded by the transcendental form of the metric; it is controlled by the polynomial condition \eqref{tripleroot}, supplemented by $A_{\rm ext}>0$. For the benchmark $M=1$, $Q=0.5$, the positive solution of Eq.~\eqref{tripleroot} that yields $A>0$ lies in a positive-$l$ sector and hence outside the globally weak energy condition (WEC) satisfying branch identified below. This provides a useful distinction between mathematical degeneracy and matter-sector viability.

The horizon analysis therefore supplies more than a numerical classification of roots. Equations~\eqref{Aext}--\eqref{lext} turn the transcendental horizon problem into an analytic critical curve, while Eq.~\eqref{tripleroot} isolates the exceptional higher-order degeneracy. This separation is useful in the remainder of the paper: optical, thermodynamic and ringdown observables must be evaluated inside the horizon-containing domain rather than over the bare $(A,l)$ plane indiscriminately.
}

\subsection{Curvature and matter-sector viability}

{
Existence of a horizon does not by itself establish the physical character of the solution. We therefore examine separately the invariant curvature and the pointwise energy conditions of the source. This is particularly important for matter-supported black holes, for which causal admissibility and material admissibility need not select the same parameter domain \cite{Visser:1992qh}.

The Yukawa contribution is finite range, but it does not regularize the center. A convenient invariant diagnostic is the Ricci scalar. Since the Maxwell stress tensor is traceless, the Ricci scalar is generated entirely by the additional anisotropic sector and is
\begin{equation}
R=
\frac{Al(Ar-4)}{4r^2}e^{-Ar/2}.
\label{RicciScalar}
\end{equation}
For $l\neq0$,
\begin{equation}
R\sim-\frac{Al}{r^2},
\qquad r\rightarrow0,
\end{equation}
so the contribution itself introduces a Ricci singularity at the origin. The Kretschmann scalar may be written compactly for any metric of the form \eqref{fmetric} as
\begin{equation}
K=
\left[f''(r)\right]^2
+\frac{4[f'(r)]^2}{r^2}
+\frac{4[1-f(r)]^2}{r^4}.
\label{Kgeneral}
\end{equation}
For $Q\neq0$, its leading central behavior remains
\begin{equation}
K\sim\frac{56Q^4}{r^8},
\qquad r\rightarrow0,
\end{equation}
as in Reissner--Nordstr\"om (RN). Thus the finite-range matter sector changes the subleading singular structure but does not soften the dominant charged singularity. The solution should consequently be interpreted as a singular charged black hole endowed with screened matter hair.
}

\medskip
\subsection{Effective stresses and energy conditions.}

{
Since the Maxwell stress tensor is conserved on its own, the additional sector is conserved separately, and its radial conservation equation gives, after using $\rho_h+p_{r,h}=0$,
\begin{equation}
p'_{r,h}+\frac{2}{r}(p_{r,h}-p_{\perp,h})=0.
\end{equation}
Hence
\begin{equation}
p_{\perp,h}
=
p_{r,h}+\frac{r}{2}p'_{r,h}
=
-\frac{A^2l}{64\pi r}e^{-Ar/2}.
\end{equation}
The total density and pressures are therefore
\begin{align}
\rho&=
\frac{Q^2}{8\pi r^4}
-\frac{Al}{16\pi r^2}e^{-Ar/2},
\label{finaldensity}\\
p_r&=-\rho,\\
p_\perp&=
\frac{Q^2}{8\pi r^4}
-\frac{A^2l}{64\pi r}e^{-Ar/2}.
\label{ptotal}
\end{align}

These expressions permit a complete analytic discussion of the standard pointwise energy conditions. First,
\begin{equation}
\rho+p_r=0,
\end{equation}
so the radial null-energy condition is identically saturated. In the tangential direction,
\begin{equation}
\rho+p_\perp=
\frac{Q^2}{4\pi r^4}
-\frac{Al(4+Ar)}{64\pi r^2}e^{-Ar/2}.
\label{NECt}
\end{equation}
For $A>0$ and $l<0$, both terms in Eqs.~\eqref{finaldensity} and \eqref{NECt} are non-negative. Consequently, the null (NEC) and weak energy conditions hold for all $r>0$ on the negative-$l$ branch.

The strong energy condition (SEC) additionally requires
\begin{equation}
\rho+p_r+2p_\perp=2p_\perp\ge0.
\end{equation}
For $l<0$, Eq.~\eqref{ptotal} is positive, so the SEC also holds globally outside the singular center.

The dominant energy condition (DEC) is more restrictive. Since $p_r=-\rho$, its radial part is saturated, while the tangential part requires both $\rho+p_\perp\ge0$ and
\begin{equation}
\rho-p_\perp=
\frac{Al(Ar-4)}{64\pi r^2}e^{-Ar/2}\ge0.
\label{DECdifference}
\end{equation}
For $l<0$, Eq.~\eqref{DECdifference} is satisfied only for
\begin{equation}
Ar\le4.
\end{equation}
Thus the WEC, NEC and SEC can hold globally on the negative-$l$ branch, whereas the DEC is generally restricted to the finite radial domain $r\le4/A$. This distinction is physically relevant and prevents positivity of the density alone from being mistaken for complete matter viability.

\begin{figure*}
\begin{center}
\subfigure{\includegraphics[width=0.45\textwidth]{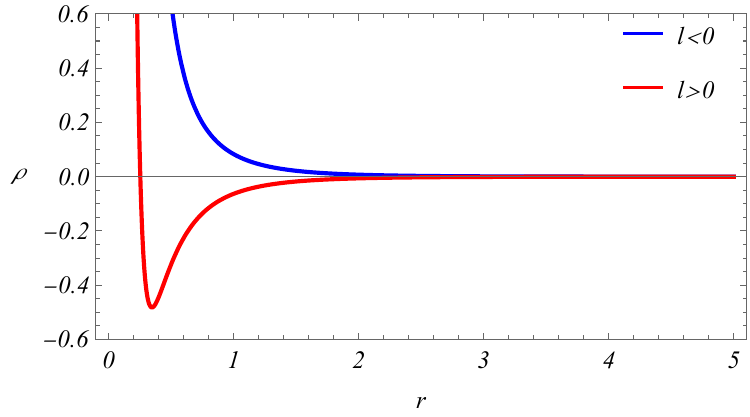}}~~~
\subfigure{\includegraphics[width=0.45\textwidth]{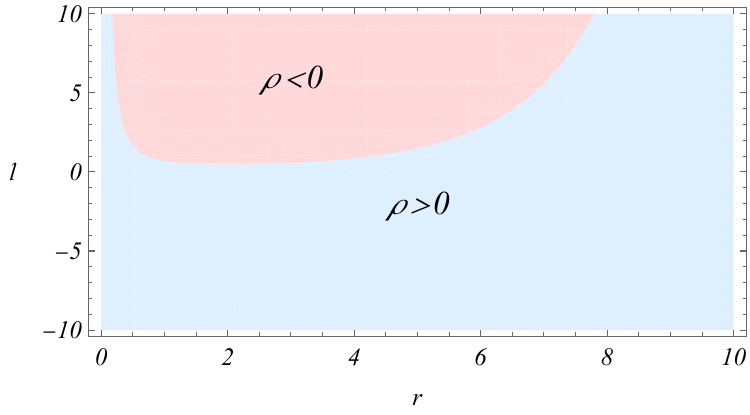}}
\end{center}
\caption{Effective density of the combined Maxwell--Yukawa source. The plots illustrate the sign of $\rho$ for the representative parameter choices indicated in the text; the complete energy-condition content is given analytically by Eqs.~\eqref{NECt}--\eqref{DECdifference}.}
\label{newfig2}
\end{figure*}

The physically distinguished sector for the subsequent black hole analysis is therefore the intersection of the horizon-containing region with $A>0$ and $l<0$. This choice guarantees asymptotic screening and global satisfaction of the WEC, NEC and SEC, while the DEC has the radial limitation stated above.

Taken together, the results of this section define the precise domain of the model used below. The geometry is generated by a nonlocal anisotropic sector rather than postulated phenomenologically; its additional mass contribution is exponentially screened and carries no independent asymptotic charge; its horizon boundary is known analytically despite the transcendental metric function; and the negative-$l$ branch simultaneously selects positive energy density and global NEC/WEC/SEC satisfaction.
}

\section{Horizon thermodynamics and local stability of the screened sector}
\label{sec:thermo}

{
The finite-range contribution affects not only the causal structure but also the
horizon mechanics. This point requires some care because the parameters $l$ and
$A$ do not have the same status as the Maxwell charge. For every finite $A>0$,
$l$ is exponentially screened from the asymptotic charge data, whereas $A$ sets
an inverse length associated with the matter profile. We therefore distinguish
between the standard thermodynamic variables associated with asymptotic charges
and an \emph{extended solution-space variation} in which the parameters of the
screened sector are also allowed to vary. This distinction is particularly
important here because the matter sector has been specified through an effective
nonlocal closure rather than through a microscopic covariant matter action.

\subsection{Horizon energy, first law and screened-sector conjugates}

Let $r_+$ denote the outer event horizon. Solving $f(r_+)=0$ for the ADM mass
gives
\begin{equation}
M(r_+,Q,l,A)
=
\frac{r_+}{2}
+\frac{Q^2}{2r_+}
-\frac{l}{2}e^{-Ar_+/2}.
\label{Mthermo}
\end{equation}
The first two terms reproduce the RN horizon energy,
whereas the last term is the finite-range contribution evaluated directly at
the horizon. Thus the same exponential that removes the Yukawa sector from the
asymptotic charge algebra remains present in the horizon energy balance.

For Einstein gravity the entropy is one quarter of the horizon area, while the
temperature is fixed by the surface gravity \cite{Bekenstein:1973ur,
Bardeen:1973gs,Hawking:1976de,Iyer:1994ys}. Hence
\begin{align}
T&=\frac{f'(r_+)}{4\pi}
=
\frac{1}{4\pi r_+}
\left[
1-\frac{Q^2}{r_+^2}
+\frac{Al}{2}e^{-Ar_+/2}
\right],
\label{TH}\\
S&=\pi r_+^2.
\label{entropyBH}
\end{align}
At fixed $(Q,l,A)$ one immediately verifies
\begin{equation}
\left(\frac{\partial M}{\partial S}\right)_{Q,l,A}=T,
\end{equation}
so the usual horizon first law is recovered on every fixed-$(l,A)$ slice of
the family.

It is also useful to enlarge the space of stationary solutions and allow the
parameters $(l,A)$ to vary. Differentiating Eq.~\eqref{Mthermo} then gives
\begin{equation}
dM=T\,dS+\Phi\,dQ+\Psi_l\,dl+\Psi_A\,dA,
\label{firstlaw}
\end{equation}
where
\begin{align}
\Phi&=\frac{Q}{r_+},\\
\Psi_l&=-\frac{1}{2}e^{-Ar_+/2},
\label{Psil}\\
\Psi_A&=\frac{lr_+}{4}e^{-Ar_+/2}.
\label{PsiA}
\end{align}
The two additional conjugates have a direct physical interpretation.
$\Psi_l$ measures the response of the horizon energy to a change in the
amplitude of the screened sector, whereas $\Psi_A$ measures the response to a
change in its inverse range. They are not independent structures generated by
two unrelated kinds of hair; rather,
\begin{equation}
\Psi_A=-\frac{lr_+}{2}\Psi_l.
\label{PsiRelation}
\end{equation}
Equation~\eqref{PsiRelation} makes explicit that both responses originate from
the same Yukawa profile.

We emphasize the status of Eq.~\eqref{firstlaw}. The variation of $Q$ is the
usual variation of a conserved Maxwell charge. By contrast, within the
effective construction adopted here, $l$ and $A$ parametrize neighboring
solutions of the nonlocal matter family. Their variations therefore define an
extended solution-space first law; they should not be interpreted, without an
underlying covariant matter action, as new Noether or Gauss-law charges.
Extended black hole thermodynamics in which dimensionful couplings or
parameters are promoted to thermodynamic variables has a well-developed
covariant formulation \cite{Iyer:1994ys,Xiao:2023lap,Hajian:2023cwe}. Our
result is consistent with that general philosophy at the level of the exact
mass function, but the present effective matter model does not by itself
supply the additional microscopic structure required to identify $A$ or $l$
with conserved coupling charges.

The integrated relation follows independently from dimensional scaling. In
geometrized units,
\begin{equation}
[S]=L^2,\qquad [Q]=[l]=L,\qquad [A]=L^{-1},
\qquad [M]=L.
\label{thermoscaling}
\end{equation}
Under
\begin{equation}
S\rightarrow\lambda^2S,\qquad
Q\rightarrow\lambda Q,\qquad
l\rightarrow\lambda l,\qquad
A\rightarrow\lambda^{-1}A,
\end{equation}
the mass obeys $M\rightarrow\lambda M$. Euler's theorem therefore yields
\begin{equation}
M=2TS+\Phi Q+\Psi_l l-\Psi_A A.
\label{Smarr}
\end{equation}
This is the screened-sector extension of the standard Smarr scaling relation
\cite{Smarr:1973zz}. Direct substitution of
Eqs.~\eqref{Mthermo}--\eqref{PsiA} verifies Eq.~\eqref{Smarr} identically.
The sign of the $A$ contribution is fixed by the negative scaling weight of
the inverse screening length.

\subsection{Thermal extremality and the geometric extremal curve}

The temperature provides a nontrivial consistency test of the causal analysis
of Sec.~\ref{sec:yukawa}. From Eq.~\eqref{TH}, thermal extremality is
\begin{equation}
T=0
\quad\Longleftrightarrow\quad
r_+^2-Q^2
+\frac{Al}{2}r_+^2e^{-Ar_+/2}=0.
\label{thermalext}
\end{equation}
On the other hand, the horizon equation can be written as
\begin{equation}
2Mr_+=r_+^2+Q^2-lr_+e^{-Ar_+/2}.
\label{horizonthermo}
\end{equation}
Eliminating $l e^{-Ar_+/2}$ between
Eqs.~\eqref{thermalext} and \eqref{horizonthermo}, and identifying the
degenerate horizon with $r_+=r_\star$, gives
\begin{equation}
A=
\frac{2(Q^2-r_\star^2)}
{r_\star(r_\star^2-2Mr_\star+Q^2)}
=A_{\rm ext}(r_\star),
\label{AextThermo}
\end{equation}
while back substitution yields
\begin{equation}
l=
\frac{r_\star^2-2Mr_\star+Q^2}{r_\star}
\exp\!\left[\frac{A_{\rm ext}(r_\star)r_\star}{2}\right]
=l_{\rm ext}(r_\star).
\label{lextThermo}
\end{equation}
Thus the zero-temperature boundary is \emph{exactly} the double-root curve
derived geometrically in Eqs.~\eqref{Aext}--\eqref{lext}. Thermal and causal
extremality are not merely numerically correlated in this family: they are two
representations of the same critical locus.

The RN limit is recovered in the expected way. When the screened contribution
vanishes at the horizon, either because $l\to0$ or because $Ar_+\gg1$, the
temperature becomes
\begin{equation}
T_{\rm RN}
=
\frac{1}{4\pi r_+}
\left(1-\frac{Q^2}{r_+^2}\right),
\end{equation}
and extremality reduces to $r_+=|Q|$. In contrast, the formal long-range limit
$A\to0$ eliminates the explicit Yukawa contribution from the temperature at
fixed $r_+$, while the horizon equation retains the shifted mass
$\widetilde M=M+l/2$. This again shows why the $A\to0$ limit should not be
identified with finite-range screened hair.

\subsection{Heat capacity and local thermodynamic stability}

The temperature contains more information than the extremal endpoint. At
fixed $(Q,l,A)$ the relevant canonical response function is
\begin{equation}
C_{Q,l,A}
\equiv
T\left(\frac{\partial S}{\partial T}\right)_{Q,l,A}
=
T\,
\frac{dS/dr_+}{dT/dr_+}.
\label{CQdef}
\end{equation}
Using Eqs.~\eqref{TH} and \eqref{entropyBH}, one obtains the closed expression
\begin{equation}
C_{Q,l,A}
=
2\pi r_+^2
\frac{
r_+^2-Q^2
+\dfrac{Al}{2}r_+^2e^{-Ar_+/2}
}{
-r_+^2+3Q^2
-\dfrac{Al}{2}r_+^2e^{-Ar_+/2}
\left(1+\dfrac{Ar_+}{2}\right)
}.
\label{CQexact}
\end{equation}
The numerator of Eq.~\eqref{CQexact} is proportional to the numerator of the
temperature. Consequently,
\begin{equation}
T=0\quad\Longrightarrow\quad C_{Q,l,A}=0
\label{Cextzero}
\end{equation}
for a generic extremal point for which the denominator remains finite. The
second distinguished locus is instead determined by
\begin{equation}
-r_+^2+3Q^2
-\frac{Al}{2}r_+^2e^{-Ar_+/2}
\left(1+\frac{Ar_+}{2}\right)=0,
\label{DaviesYukawa}
\end{equation}
where $dT/dr_+=0$ and the heat capacity diverges. This is the finite-range
generalization of the familiar fixed-charge heat-capacity divergence of the
RN black hole \cite{Davies:1977bgr}. In the RN limit,
Eq.~\eqref{DaviesYukawa} reduces to
\begin{equation}
r_D=\sqrt{3}\,|Q|.
\label{DaviesRN}
\end{equation}

The terminology associated with Eq.~\eqref{DaviesYukawa} deserves care. A pole
of a response function is sufficient to identify a boundary of local thermal
stability in the specified ensemble, but by itself it does not establish a
global thermodynamic phase transition. Such a statement would require the
appropriate thermodynamic potential and a comparison of competing equilibrium
branches. We therefore refer to Eq.~\eqref{DaviesYukawa} as the
\emph{heat-capacity divergence} or \emph{local-stability boundary}.

For positive temperature, the sign of $C_{Q,l,A}$ is determined by the
denominator of Eq.~\eqref{CQexact}. On the physical $A>0$, $l<0$ branch the
Yukawa contribution shifts both the extremal radius and the local-stability
boundary away from their RN values. The resulting interval between the
zero-temperature endpoint and the first heat-capacity pole can support
$C_{Q,l,A}>0$, whereas sufficiently large outer horizons recover the familiar
negative-capacity behavior because
\begin{equation}
C_{Q,l,A}\sim -2\pi r_+^2,
\qquad
r_+\rightarrow\infty.
\end{equation}
Strong screening drives both characteristic radii back toward their RN
limits. The same finite-range mechanism that suppresses the contribution in
the geometry therefore also suppresses it in the local thermodynamic response.

\subsection{Horizon-local hair and the role of the screening gradient}

The thermodynamic expressions become especially transparent when written in
terms of quantities evaluated at the horizon. Define
\begin{equation}
\chi_+
\equiv
\frac{l}{r_+}e^{-Ar_+/2},
\qquad
x_+\equiv Ar_+,
\qquad
q_+\equiv\frac{Q}{r_+}.
\label{horizonhair}
\end{equation}
Here $\chi_+$ measures the residual amplitude of the screened hair at the
horizon, while $x_+$ measures how rapidly the Yukawa profile varies on the
horizon scale. The dimensionless temperature becomes
\begin{equation}
4\pi r_+T
=
1-q_+^2+\frac{x_+\chi_+}{2},
\label{Tlocalhair}
\end{equation}
and the local-stability boundary takes the compact form
\begin{equation}
-1+3q_+^2
-\frac{x_+\chi_+}{2}
\left(1+\frac{x_+}{2}\right)=0.
\label{DaviesLocalHair}
\end{equation}

Equations~\eqref{Tlocalhair} and \eqref{DaviesLocalHair} show that the
thermodynamics cannot be organized by $\chi_+$ alone. The amount of hair that
survives at the horizon is important, but the response functions also retain
independent sensitivity to the screening gradient through $x_+=Ar_+$. This is
the horizon analogue of the structure found later in the optical sector:
photon trapping is organized by the residual hair at the photon sphere but
also probes the radial gradient of the Yukawa profile. The common lesson is
that the bare parameter $l$ is not by itself a measure of strong-field
contribution. What matters is the residual hair at the radial scale sampled by
the observable together with the local screening rate.

This perspective unifies the limiting regimes. For $x_+\gg1$,
$\chi_+\rightarrow0$ exponentially and the thermodynamic response approaches
RN even when the bare value of $|l|$ is appreciable. For $x_+\lesssim1$, the
screened sector reaches the horizon and can substantially shift both
extremality and the heat-capacity pole. The limit $A\to0$ is again exceptional:
the range becomes infinite and $l$ is absorbed into the effective mass rather
than behaving as finite-range hair.

Figure~\ref{fig:thermo_response} summarizes the thermodynamic effect of the
finite-range sector for the benchmark $M=1$ and $Q=0.5$. The left panel shows
the Hawking temperature as a function of the outer-horizon radius. Besides the
RN curve, representative configurations on the physically distinguished
$A>0$, $l<0$ branch are displayed. The zero of each curve identifies the
corresponding extremal endpoint and makes explicit the displacement of thermal
extremality induced by the screened sector. The right panel shows the
fixed-$(Q,l,A)$ heat capacity. Its poles are the roots of
Eq.~\eqref{DaviesYukawa} and therefore identify the local-stability boundaries
discussed above. The simultaneous displacement of the zero-temperature radius
and the heat-capacity pole demonstrates that finite-range hair modifies two
distinct thermodynamic scales rather than producing only a rigid shift of the
RN response.

\begin{figure*}[t]
\centering
\includegraphics[width=0.92\textwidth]{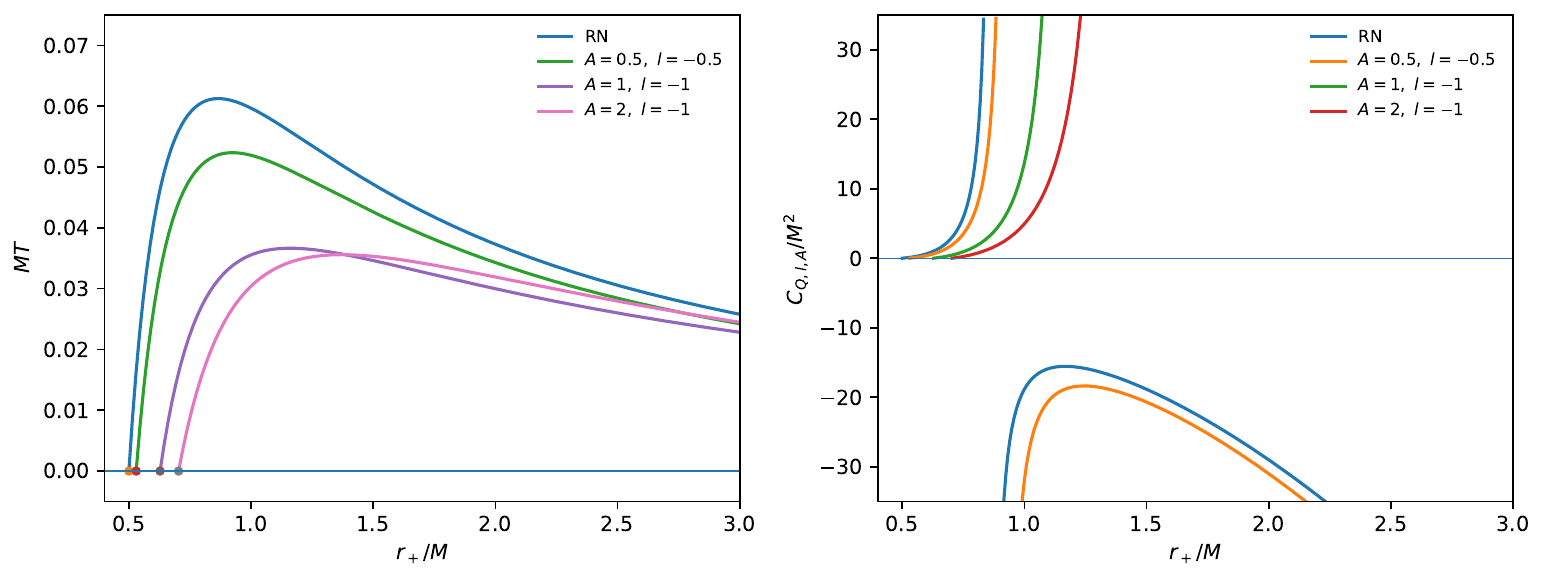}
\caption{{Thermodynamic response of the screened charged black
hole for $M=1$ and $Q=0.5$. Left: Hawking temperature as a function of the
outer-horizon radius for RN and representative configurations on the
$A>0$, $l<0$ branch. The zeros locate the corresponding extremal endpoints.
Right: heat capacity at fixed $(Q,l,A)$. The divergences occur at the roots of
Eq.~\eqref{DaviesYukawa} and separate branches with opposite local thermal
response. They are interpreted here as local-stability boundaries in the
specified ensemble, not by themselves as evidence for a global phase
transition. As the Yukawa contribution becomes increasingly screened, both
the thermal extremality and the local response approach their RN behavior.}}
\label{fig:thermo_response}
\end{figure*}

The thermodynamic sector therefore provides an independent strong-field
diagnostic of the same screening mechanism that governs the causal and optical
properties of the solution. The geometric extremal curve reappears exactly as
the zero-temperature locus; the amplitude and range of the matter contribution
enter a consistent extended solution-space first law and Smarr identity; and
the heat capacity introduces a second critical curve that separates locally
stable and unstable thermal response at fixed $(Q,l,A)$. Most importantly,
these effects are controlled by the contribution that survives at the horizon,
not simply by the asymptotic input amplitude. This gives the thermodynamic
analysis a direct conceptual connection with the photon-sphere and ringdown
sectors studied below.
}

\section{Optical signatures of the finite-range Yukawa contribution}
\label{sec:optics}

{
The optical sector probes precisely the radial domain in which the finite-range
hair survives. For every finite $A>0$ the geometry is asymptotically
RN, so an optical departure from RN cannot be attributed
to a change of the ADM mass. It must instead be generated by the residual
Yukawa contribution in the strong-field region. This makes the unstable null
orbit especially useful: it converts the global parameters $(A,l)$ into a
local measure of how much screened hair actually reaches the photon region.
}

\subsection{Photon sphere}

{
For null equatorial geodesics the conserved quantities are
$E=f\dot t$ and $L=r^2\dot\phi$, and the radial equation can be written as
\begin{equation}
 \dot r^2+V_{\rm eff}(r)=E^2,
 \qquad
 V_{\rm eff}(r)=\frac{L^2}{r^2}f(r).
 \label{veffeq}
\end{equation}
The unstable circular orbit is determined by
\begin{equation}
 2f(r_{\rm ph})-r_{\rm ph}f'(r_{\rm ph})=0.
 \label{photoncondition}
\end{equation}
For the exact metric~\eqref{fmetric}, this condition becomes
\begin{equation}
 2-\frac{6M}{r_{\rm ph}}+\frac{4Q^2}{r_{\rm ph}^2}
 -l e^{-A r_{\rm ph}/2}
 \left(\frac{3}{r_{\rm ph}}+\frac{A}{2}\right)=0.
 \label{photonYukawa}
\end{equation}
Equation~\eqref{photonYukawa} already exposes the essential finite-range
physics. The contribution is not controlled by the bare amplitude $l$ alone:
the exponential determines how much of that amplitude survives at the null
orbit, while the factor $3/r_{\rm ph}+A/2$ shows that the orbit also probes the
radial gradient of the Yukawa profile. Consequently, two configurations with
the same local value of the metric contribution need not have exactly the same
photon sphere if their screening lengths differ.

The RN reference radius is
\begin{equation}
 r_0\equiv r_{\rm ph}^{\rm RN}
 =\frac{3M+\sqrt{9M^2-8Q^2}}{2}.
 \label{rphRN}
\end{equation}
A particularly transparent analytic result follows by expanding
$r_{\rm ph}=r_0+\delta r_{\rm ph}+{\cal O}(l^2)$ around RN. To first order in
the Yukawa amplitude,
\begin{equation}
 \delta r_{\rm ph}
 =
 \frac{l\,r_0^2(6+A r_0)}
 {4(3Mr_0-4Q^2)}
 e^{-A r_0/2}.
 \label{drph}
\end{equation}
This expression makes several properties immediate. On the physically distinguished
$A>0$, $l<0$ branch, $\delta r_{\rm ph}<0$: the screened sector shifts the
unstable null orbit inward relative to RN. The magnitude of the shift is
exponentially suppressed when $A r_0\gg1$, even if $|l|$ itself is not small.
Conversely, the limit $A\to0$ does not represent screened hair; it reproduces
the photon sphere of an RN geometry with the shifted mass
$M+l/2$, consistently with the asymptotic discussion in
Sec.~\ref{sec:yukawa}.

To organize the nonperturbative solutions it is useful to define the local
residual hair
\begin{equation}
 \chi_{\rm ph}\equiv
 \frac{l}{r_{\rm ph}}e^{-A r_{\rm ph}/2}.
 \label{effectivehair}
\end{equation}
Figure~\ref{fig:optical_local} shows the fractional shifts of the photon sphere
and shadow against $\chi_{\rm ph}$ for several screening lengths. Near the RN
limit the families approach a common trend, confirming that the amount of hair
surviving at the photon region is a more physical organizing variable than
$l$ alone. The collapse is not exact away from that limit. This is not a
numerical accident: Eq.~\eqref{photonYukawa} contains the independent gradient
factor $A/2$. The departure from a universal curve therefore measures the
fact that null trapping is sensitive both to the local amplitude of the hair
and to how rapidly that hair is varying across the photon region.

\begin{figure*}[t]
\centering
\includegraphics[width=0.92\textwidth]{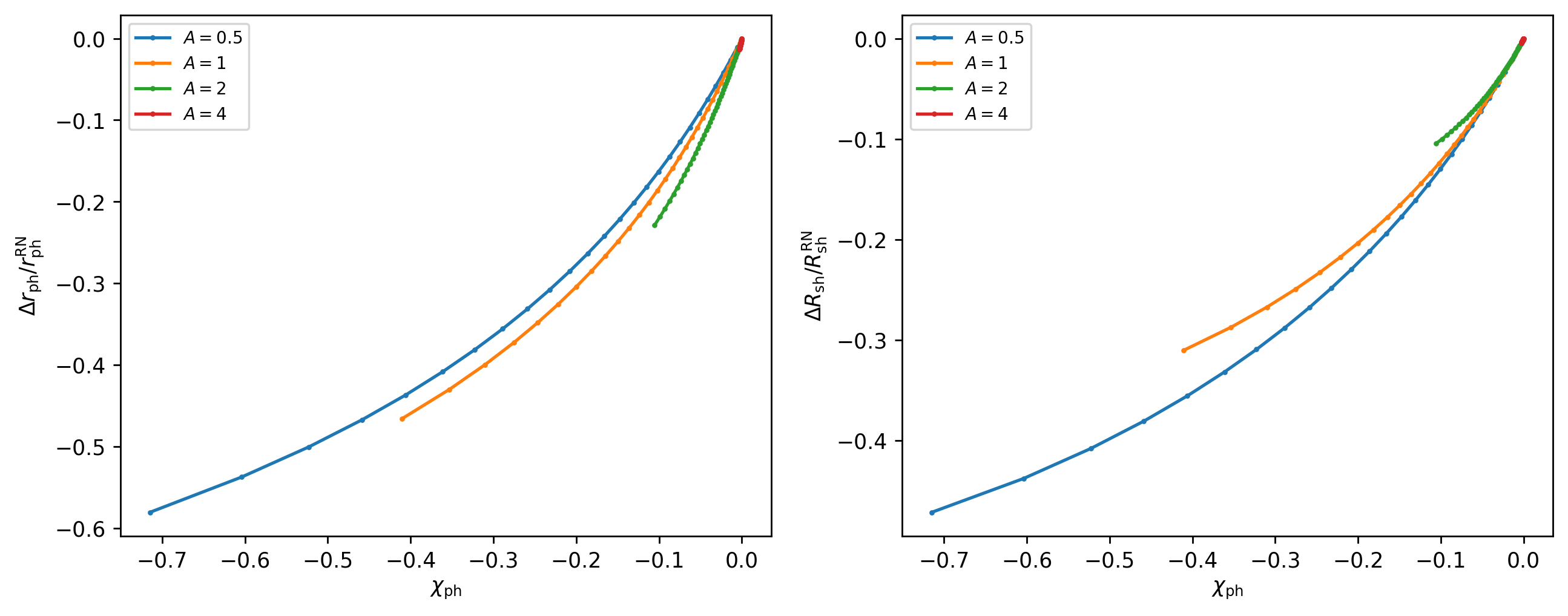}
\caption{{Local organization of the optical contribution for
$M=1$ and $Q=0.5$. Left: fractional photon-sphere shift relative to RN as a
function of the residual hair $\chi_{\rm ph}$. Right: corresponding fractional
shadow shift. Different values of $A$ approach a common RN trend when the
surviving hair is weak, but do not collapse onto an exact universal curve.
The residual separation is the optical imprint of the Yukawa gradient,
explicitly present in Eq.~\eqref{photonYukawa}.}}
\label{fig:optical_local}
\end{figure*}
}

\subsection{Shadow radius}

{
For an observer at spatial infinity the critical impact parameter is
\begin{equation}
 R_{\rm sh}\equiv b_c
 =\frac{r_{\rm ph}}{\sqrt{f(r_{\rm ph})}}.
 \label{shadowradius}
\end{equation}
Although this expression contains both $r_{\rm ph}$ and $f(r_{\rm ph})$, its
linear response around RN is simpler than might be expected. Varying
$b=r/\sqrt{f}$ gives
\begin{equation}
 \frac{\delta b}{b}
 =
 \left[
 \frac{1}{r}-\frac{f'}{2f}
 \right]_{r_0}\delta r
 -\frac{\delta f(r_0)}{2f_{\rm RN}(r_0)}.
 \label{shadowvariation}
\end{equation}
The coefficient multiplying $\delta r$ vanishes identically because the RN
photon sphere obeys $r_0f'_{\rm RN}(r_0)=2f_{\rm RN}(r_0)$. Hence the
first-order shadow shift does not require the first-order displacement of the
orbit:
\begin{equation}
 \frac{\delta R_{\rm sh}}{R_{\rm sh}^{\rm RN}}
 =
 \frac{l}{2r_0 f_{\rm RN}(r_0)}
 e^{-A r_0/2}
 +{\cal O}(l^2).
 \label{dshadow}
\end{equation}
This cancellation is useful physically. At leading order the shadow measures
the metric contribution evaluated at the unperturbed critical orbit, whereas
the photon-sphere displacement also resolves its radial derivative. Thus
$r_{\rm ph}$ and $R_{\rm sh}$ arise from the same null orbit but encode the
Yukawa contribution in slightly different combinations. For $l<0$,
Eq.~\eqref{dshadow} predicts a smaller shadow than RN, while strong screening
drives the correction exponentially to zero.

This distinction also explains why the bare magnitude of the hair is a poor
proxy for optical visibility. A configuration may possess sizeable $|l|$ and
still be almost RN-like if its screening length is short compared with the
photon-sphere scale. Figure~\ref{fig:optical_shadow} makes this statement
global. The left panel maps the full numerical fractional shadow contribution over the
$(A,l)$ plane for the benchmark $M=1$, $Q=0.5$, retaining only configurations
with an outer horizon. The rapid loss of optical sensitivity toward large $A$
exhibits an ``optically screened'' region in which nonzero strong-field hair
produces only a small change of the critical impact parameter. The blank
domain is not an observational degeneracy but the region in which the
corresponding black hole exterior is absent.

The right panel compares two different evaluations of the same observable.
The solid curves are obtained by solving the full photon-sphere equation
\eqref{photonYukawa} numerically and then evaluating
Eq.~\eqref{shadowradius}, without expanding in $l$. The dashed curves instead
use the first-order RN expansion~\eqref{dshadow}, obtained by retaining only
terms linear in $l$ around $l=0$. Their agreement therefore identifies the
domain in which the RN perturbative description is reliable. For weak residual
hair the two calculations coincide, whereas long-range and sufficiently
large-amplitude corrections require the full nonlinear result. Importantly,
increasing $|l|$ does not by itself determine the breakdown of the expansion:
configurations with large $A$ can remain close to RN because the exponential
suppression acts before the photon region is reached. The effective measure of
perturbativity is therefore the hair surviving in the strong-field region
rather than the bare amplitude alone.

\begin{figure*}[t]
\centering
\includegraphics[width=0.92\textwidth]{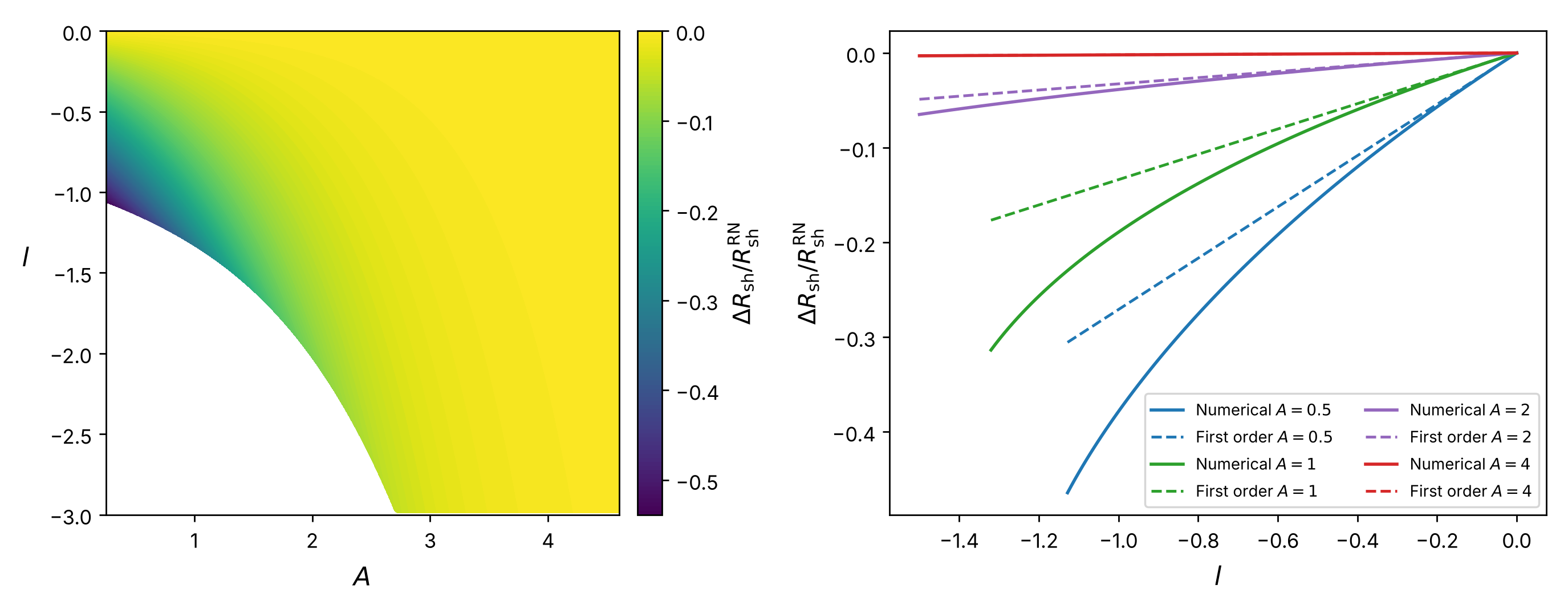}
\caption{{Global and perturbative diagnostics of the Yukawa
shadow for $M=1$ and $Q=0.5$. Left: fractional deviation
$(R_{\rm sh}-R_{\rm sh}^{\rm RN})/R_{\rm sh}^{\rm RN}$ obtained from the
full photon-sphere equation over the $(A,l)$ plane, restricted to
horizon-containing configurations. Strong screening produces a broad RN-like
optical region even at nonzero $l$. Right: the solid curves are the full
numerical result obtained from Eqs.~\eqref{photonYukawa} and
\eqref{shadowradius}, while the dashed curves are the first-order expansion
around RN given by Eq.~\eqref{dshadow}. Their separation directly measures
the breakdown of the linear-in-$l$ approximation.}}
\label{fig:optical_shadow}
\end{figure*}

The two optical diagnostics therefore expose a useful hierarchy. The shadow
size is exceptionally clean at first order because the displacement of the
critical radius cancels from its variation, whereas the photon-sphere equation
retains direct sensitivity to the Yukawa gradient. This is precisely the
structure needed for the ringdown analysis below. In the eikonal regime the
real quasinormal frequency is tied to the same critical impact parameter,
while finite-multipole potentials and their damping rates additionally probe
derivatives of the geometry. Optical and spectral observables are therefore
expected to share the same screening trend without being exactly redundant.
}

{
\subsection{Constraints from the Event Horizon Telescope}
\label{sec:eht}
The Event Horizon Telescope (EHT) has measured the angular shadow diameter of the supermassive black holes M87* and Sgr A*~\cite{EventHorizonTelescope:2019dse,EventHorizonTelescope:2019ggy,EventHorizonTelescope:2022wkp,EventHorizonTelescope:2022xqj}. Expressed in units of the source mass, these measurements give $d_{\rm sh}/M=11.0\pm1.5$ for M87* and $d_{\rm sh}/M=9.5\pm1.4$ for Sgr A*, where $d_{\rm sh}=2R_{\rm sh}$ is the shadow diameter and $R_{\rm sh}$ is the critical impact parameter of Eq.~\eqref{shadowradius}. Since the model shadow radius is computed exactly as a function of $(A,l)$ at fixed $Q/M$, these two measurements can be compared directly against the geometry developed above, without any additional modeling, in the spirit of shadow-based constraints on modified black holes~\cite{Konoplya:2019sns,Konoplya:2022hbl,Konoplya:2020ibi}.
Figure~\ref{fig:eht_overlay} shows $d_{\rm sh}/M$ as a function of $A$ for three representative values of $l$ on the physically distinguished branch, together with the $1\sigma$ EHT bands for both sources. Sufficiently large $A$ reproduces the RN value $d_{\rm sh}/M\simeq9.94$ for $Q=0.5$, which already lies inside both bands; stronger corrections (smaller $A$ at fixed $l$, or larger $|l|$) push the curve outside the Sgr A* band first, since it is the tighter of the two constraints in this unit system.
\begin{figure}[t]
\centering
\includegraphics[width=0.48\textwidth]{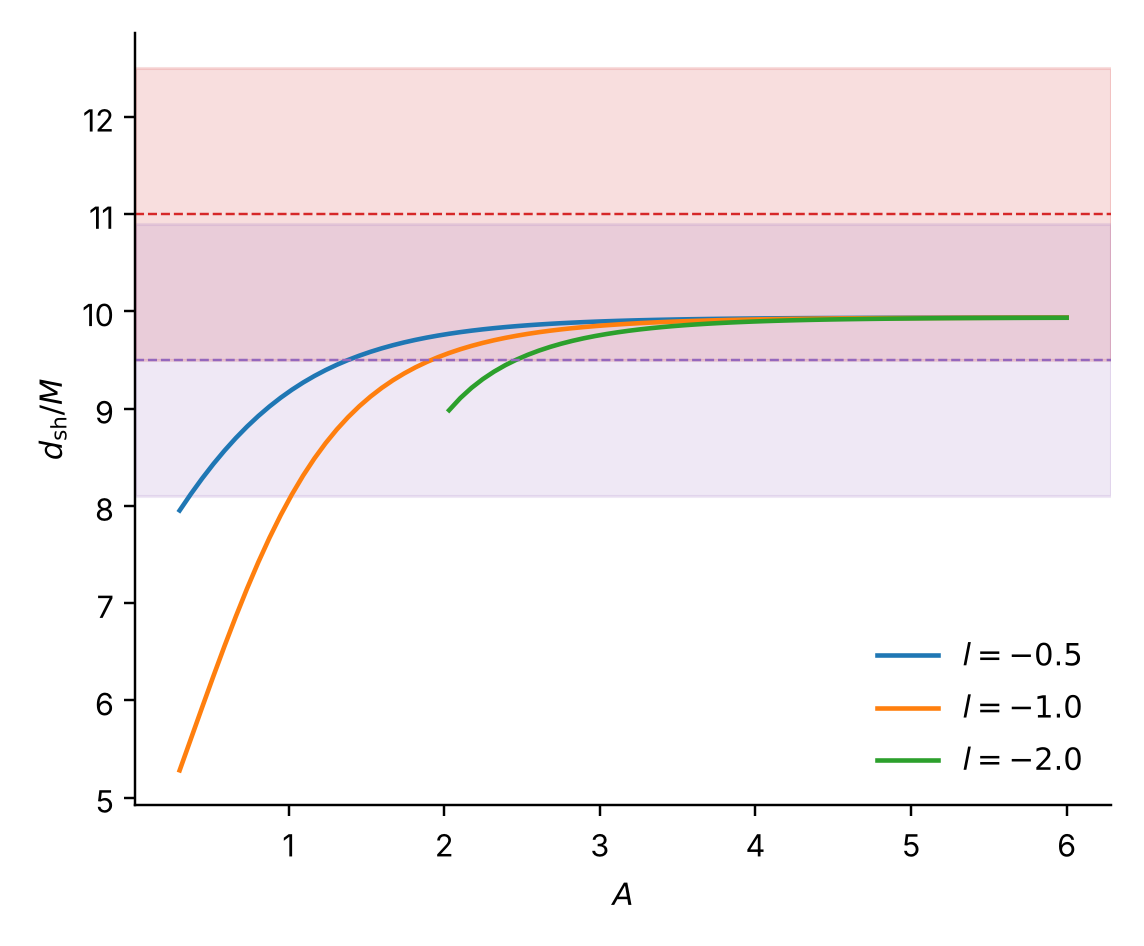}
\caption{Shadow diameter $d_{\rm sh}/M=2R_{\rm sh}/M$ as a function of $A$ for $l=-0.5,-1,-2$ at $M=1$, $Q=0.5$, compared with the $1\sigma$ EHT bands for M87* (upper, red) and Sgr A* (lower, purple).}
\label{fig:eht_overlay}
\end{figure}
A joint constraint on the full $(A,l)$ plane follows from the combined $\chi^2$ statistic
\begin{equation}
\chi^2(A,l)=\left(\frac{d_{\rm sh}(A,l)-d_{\rm M87^*}}{\sigma_{\rm M87^*}}\right)^2
+\left(\frac{d_{\rm sh}(A,l)-d_{\rm Sgr\,A^*}}{\sigma_{\rm Sgr\,A^*}}\right)^2,
\label{chi2def}
\end{equation}
evaluated on the horizon-containing, $A>0$, $l<0$ domain. Figure~\ref{fig:eht_chi2} shows the resulting $1\sigma$ and $2\sigma$ confidence regions ($\Delta\chi^2=2.30,\,6.18$). The best fit, $\chi^2_{\rm min}\simeq0.6$, is reached for $A\gtrsim5$ at any $l$ in the range scanned, i.e.\ it coincides with the RN limit of the model; the current EHT precision therefore constrains the Yukawa sector from above in $|l|/A$ but does not yet isolate a preferred finite-range configuration away from RN. This is consistent with the screening mechanism identified throughout this work: once $Ar_{\rm ph}\gg1$, the shadow becomes observationally indistinguishable from RN regardless of the bare amplitude $l$.
\begin{figure}[t]
\centering
\includegraphics[width=0.48\textwidth]{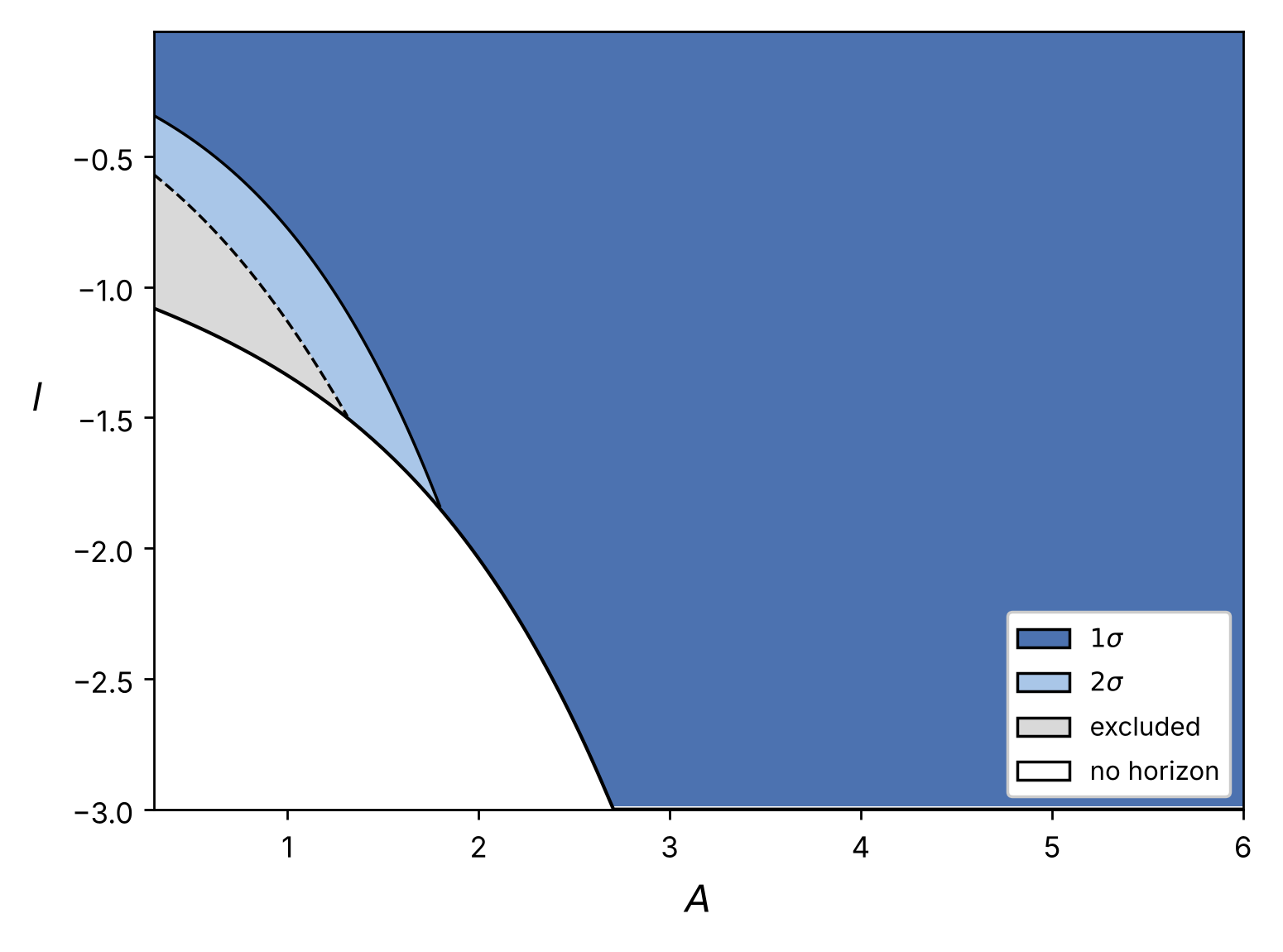}
\caption{Joint $1\sigma$ and $2\sigma$ confidence regions in the $(A,l)$ plane from Eq.~\eqref{chi2def}, combining the M87* and Sgr A* shadow-diameter measurements.}
\label{fig:eht_chi2}
\end{figure}
}

\subsection{Observational appearance}
\label{sec:obsappearance}

We consider a black hole spacetime possessing a single photon sphere, governed by the effective potential
\begin{equation}
    W(r) = \frac{f(r)}{r^2} \, ,
\end{equation}
where $f(r)$ is the metric function of Eq.~\eqref{fmetric}, so that $W(r)=V_{\rm eff}(r)/L^{2}$ is the effective potential of null geodesics per unit squared angular momentum. This potential has a single local maximum, located at the photon sphere radius $r_{\rm ph}$. A photon whose impact parameter approaches the critical value $b_c$ can wind around the photon sphere multiple times before either escaping to infinity or falling into the horizon, producing a family of self-similar photon rings labeled by the half-orbit number $n = \phi/2\pi$; in the limit $n\to\infty$ these accumulate onto the critical curve.

To construct images we place a distant observer at large $r$ and integrate null geodesics backward from the observer's screen via
\begin{equation}
    \frac{d\phi}{dr} = \pm \frac{b}{r^2}\frac{1}{\sqrt{1-b^2 W(r)}} \, ,
\end{equation}
with the sign choice tracking whether the ray is outgoing or ingoing at a given point along its trajectory. The number of times a ray crosses the equatorial plane, $n=\lfloor (\phi+\pi/2)/\pi \rfloor$, classifies it as a direct ($n=1$), lensed ($n=2$) or photon-ring ($n\geq 3$) ray.

To generate images we adopt a simplified accretion-disk model used in earlier work: geometrically thin, optically thin, and viewed face-on, with monochromatic disk-frame intensity $I_\nu^e = \tilde I(r)$ given by the Johnson Standard Unbound (SU) distribution~\cite{GLM2020,Vincent2022}
\begin{equation}\label{eq: emission}
    \tilde{I}(r;\,\gamma,\mu,\sigma)
    =
    \frac{\displaystyle
    \exp\!\left\{-\tfrac{1}{2}\left[\gamma
    +\arcsin\!\left(\tfrac{r-\mu}{\sigma}\right)\right]^{2}\right\}}
    {\sqrt{(r-\mu)^{2}+\sigma^{2}}}\, .
\end{equation}
The three emission profiles (GLM1, GLM2, GLM3) used in this work, which correspond to panels (a), (b) and (c) in the figures, are specified in \cite{Macedo:2025ipc} and \cite{Gralla2019}. The observed intensity at the detector plane is obtained by summing the contributions from each successive crossing of the ray with the disk~\cite{Luminet1979,Gralla2019}
\begin{equation}
    I_{\rm obs}(b)
    =
    \sum_{m=1}^{n}
    \,f^{2}(r)\,\tilde{I}(r)\Big|_{r\,=\,r_m(b)},
    \label{eq:Iobs}
\end{equation}
where $r_m(b)$ is the transfer function giving the disk radius hit by a ray of impact parameter $b$ at its $m$-th crossing, and $f^2(r)$ is the redshift factor for a static emitter (bolometric intensity). Here $n=3$ typically suffices for black hole images, since the contribution of each additional crossing is exponentially suppressed.

While the shadow radius provides a useful global characterization of the geometry, a more realistic description of observational signatures requires the construction of synthetic images through ray tracing.

\begin{figure}[H]
    \centering
    \includegraphics[width=0.32\linewidth]{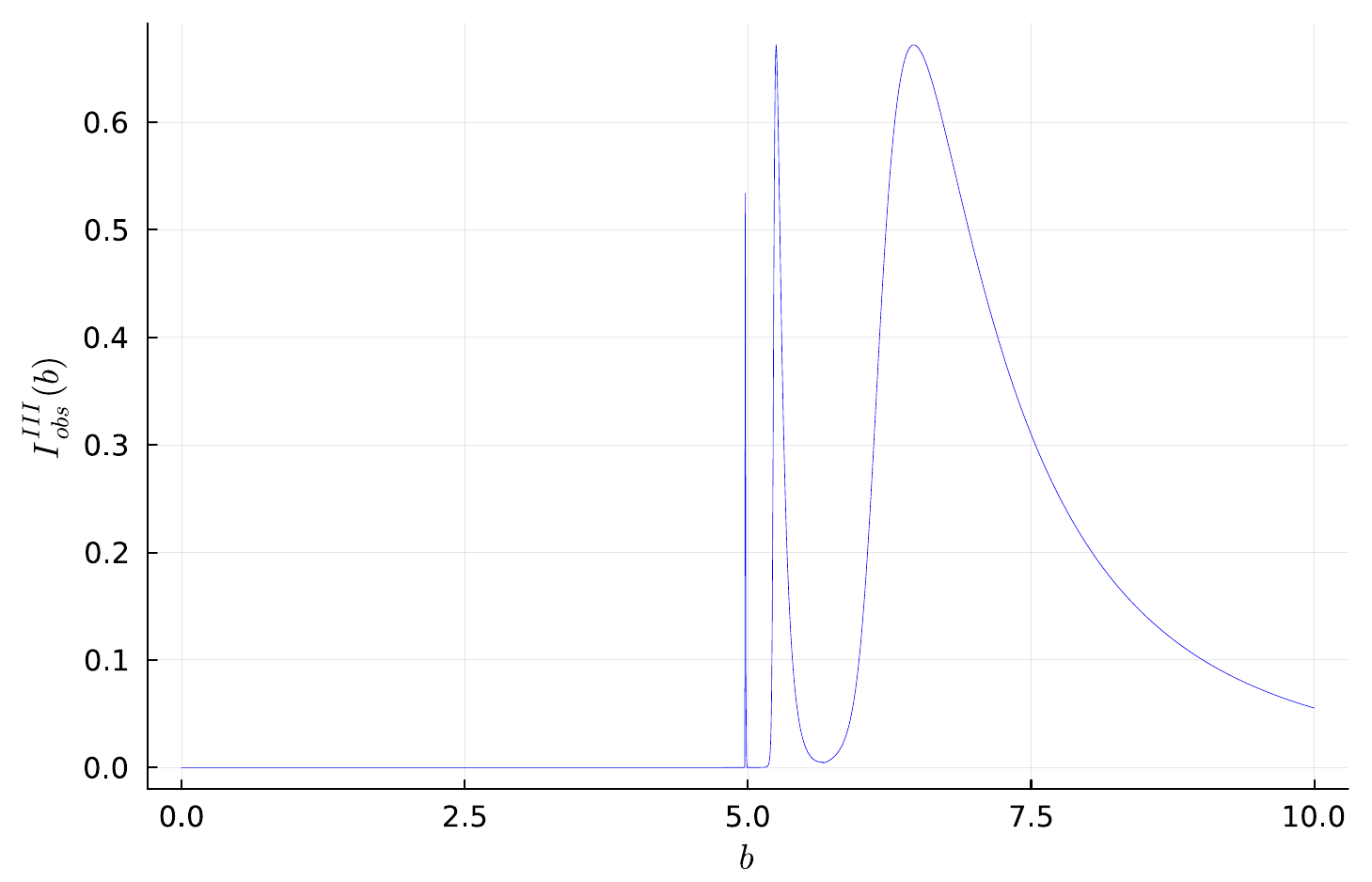}
    \includegraphics[width=0.32\linewidth]{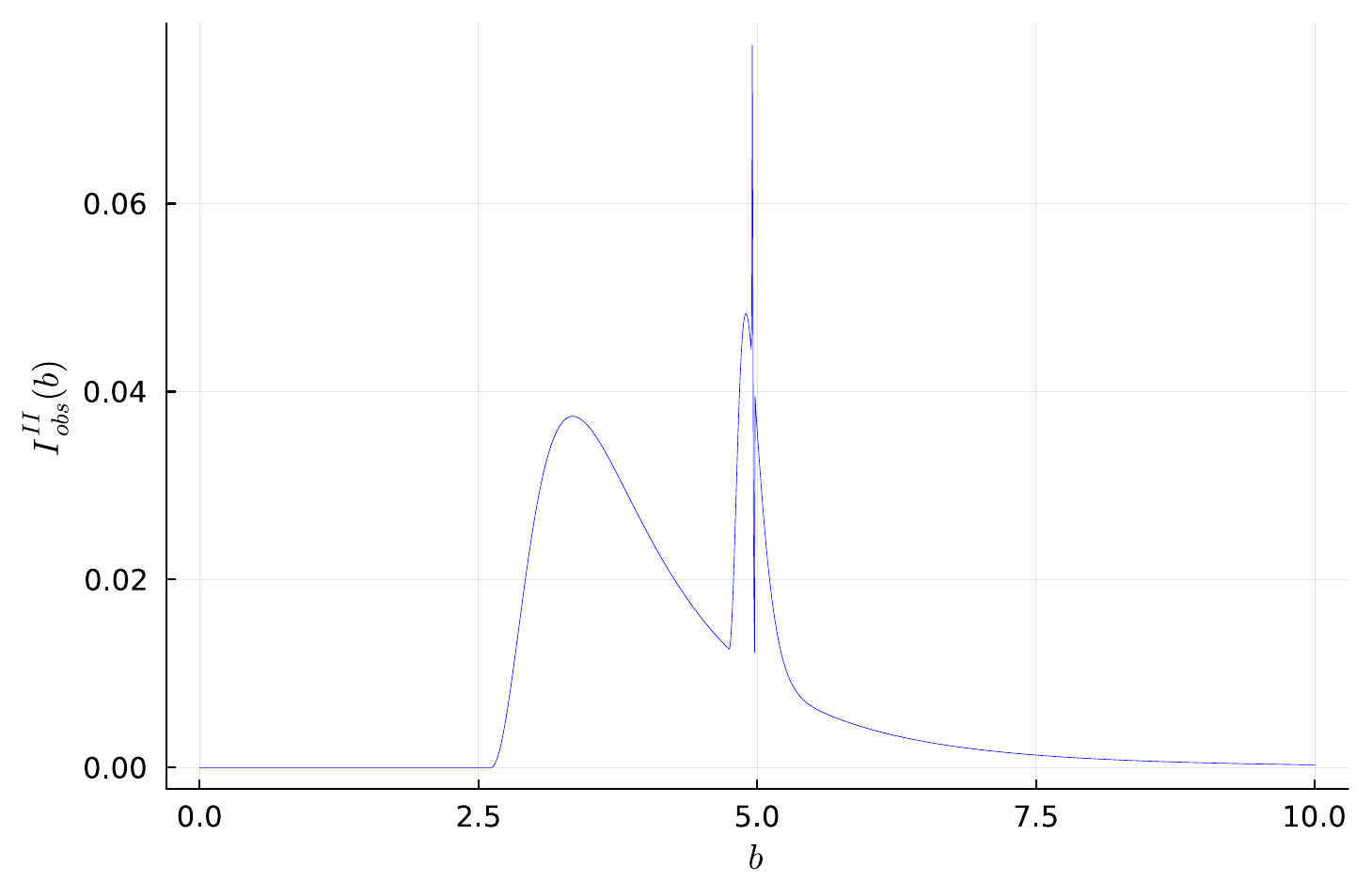}
    \includegraphics[width=0.32\linewidth]{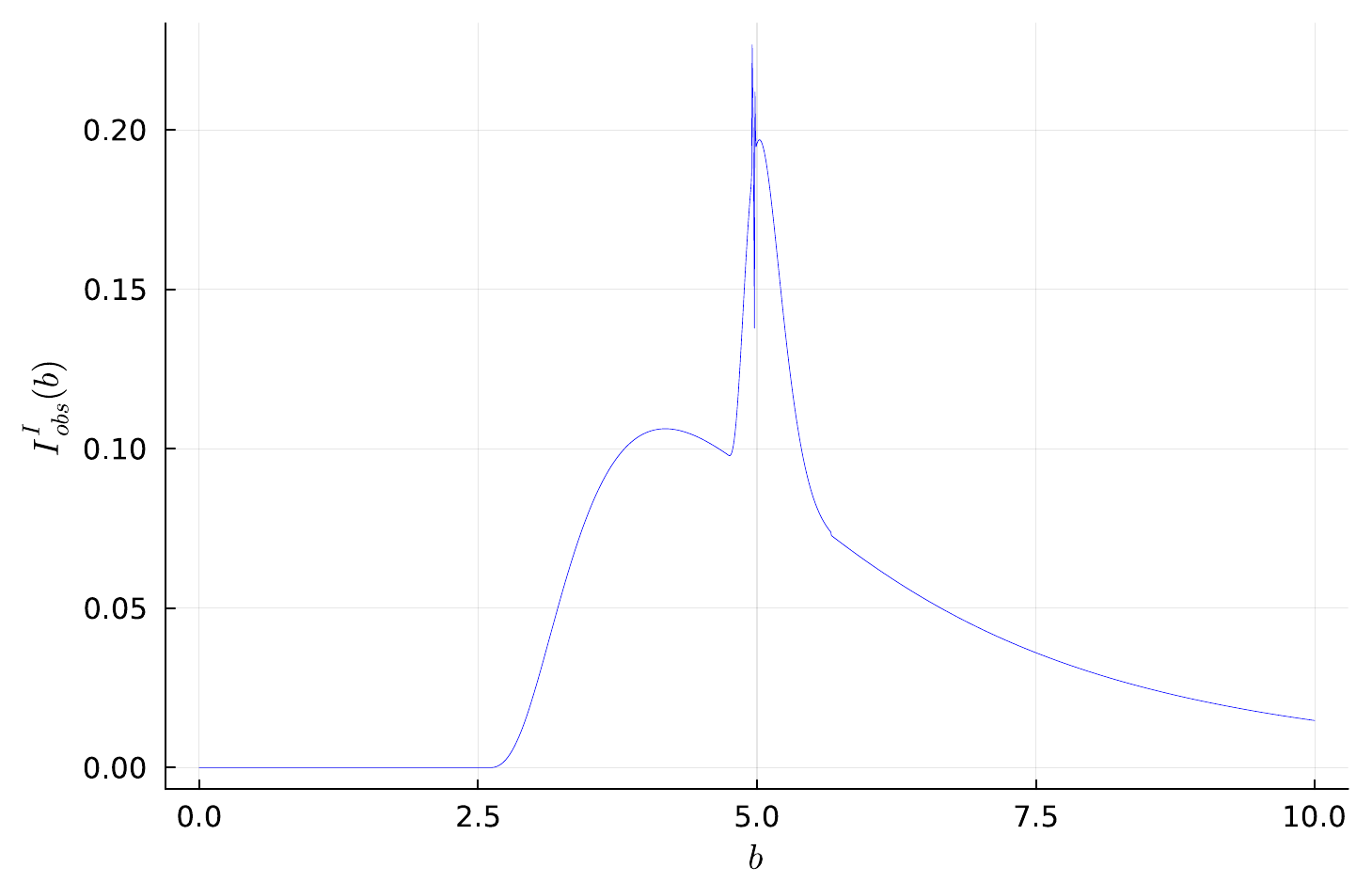}

    \includegraphics[width=0.32\linewidth]{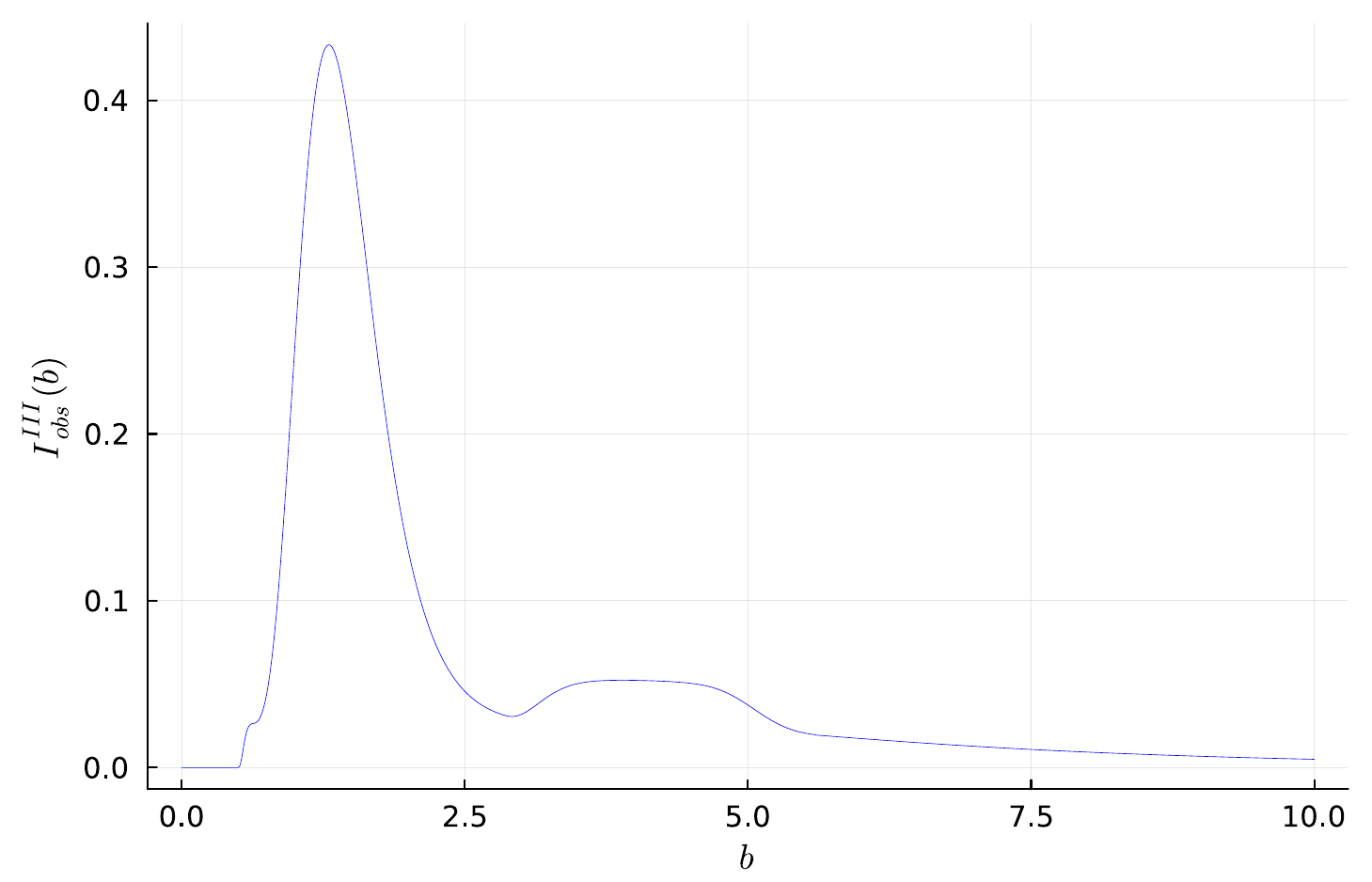}
    \includegraphics[width=0.32\linewidth]{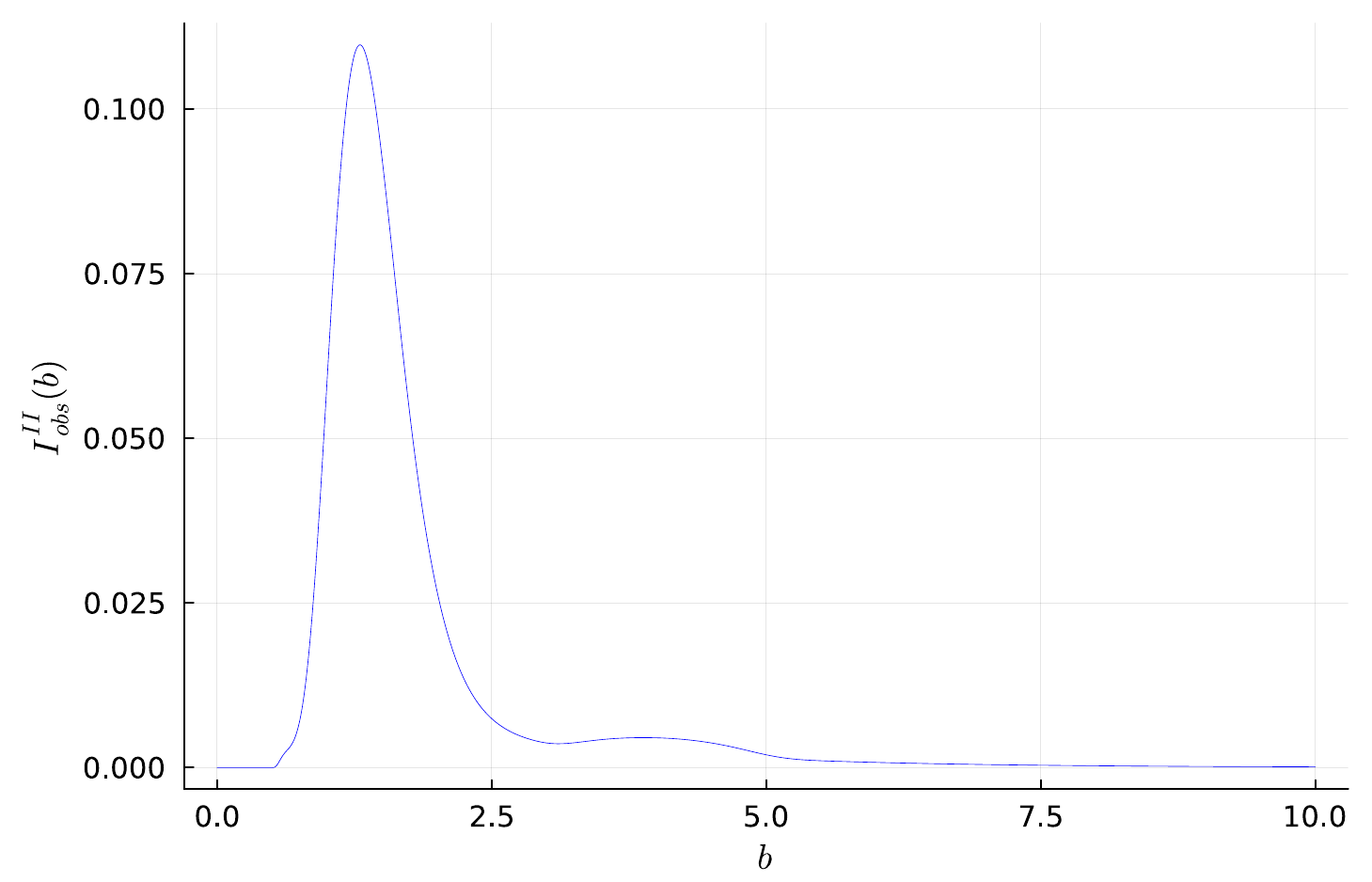}
    \includegraphics[width=0.32\linewidth]{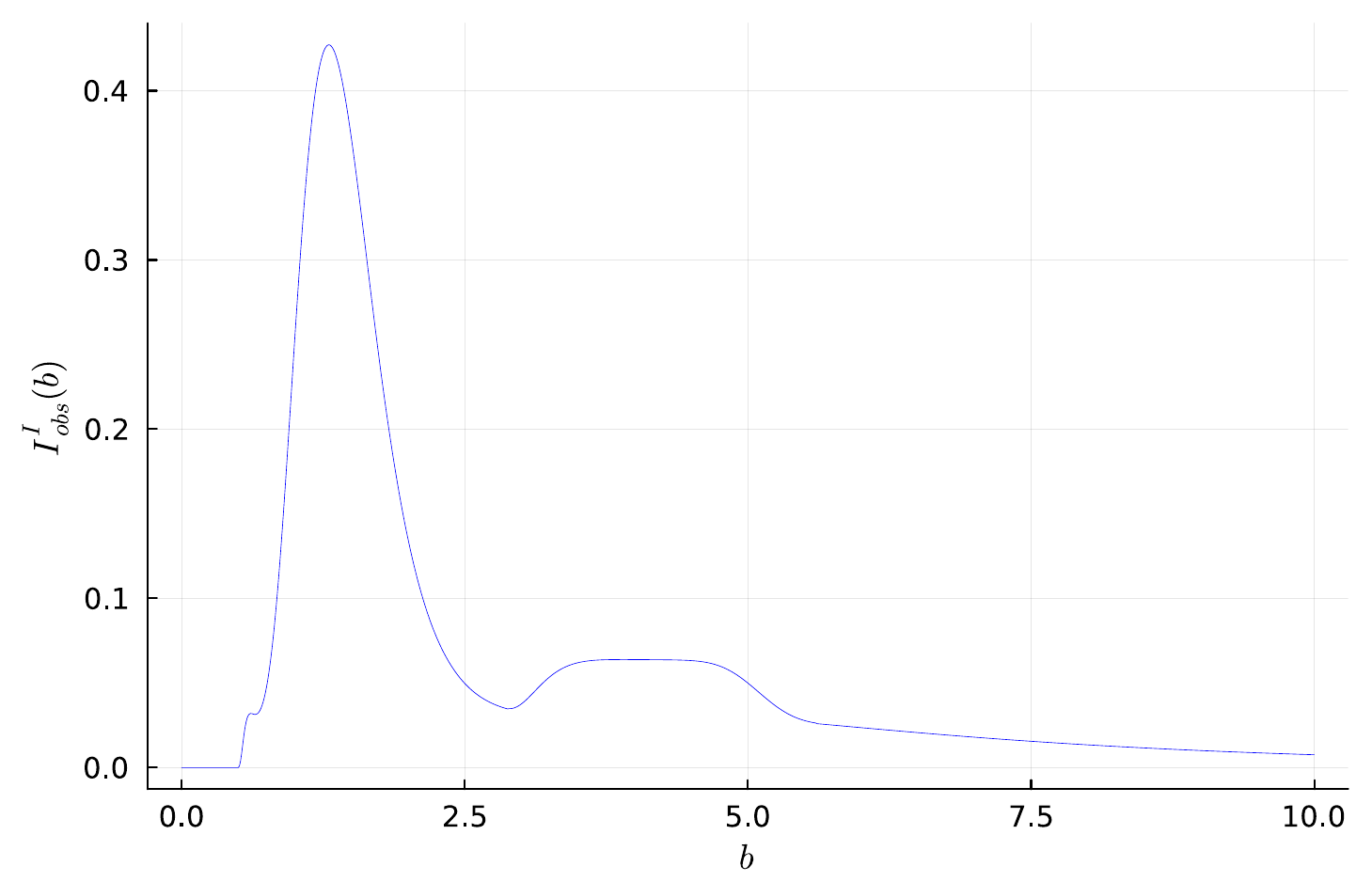}
    \caption{The top panels show the observed intensity profile for a black hole with parameters $M = 1$, $Q = 0.5$, $l = -5$ and $A = 6$. The bottom panels show the observed intensity profile for a naked singularity with parameters $M = 1$, $Q = 0.5$, $l = -10$ and $A = 3$. }
    \label{fig: absortion}
\end{figure}

\begin{figure}[H]
    \centering
    \includegraphics[width=0.32\linewidth]{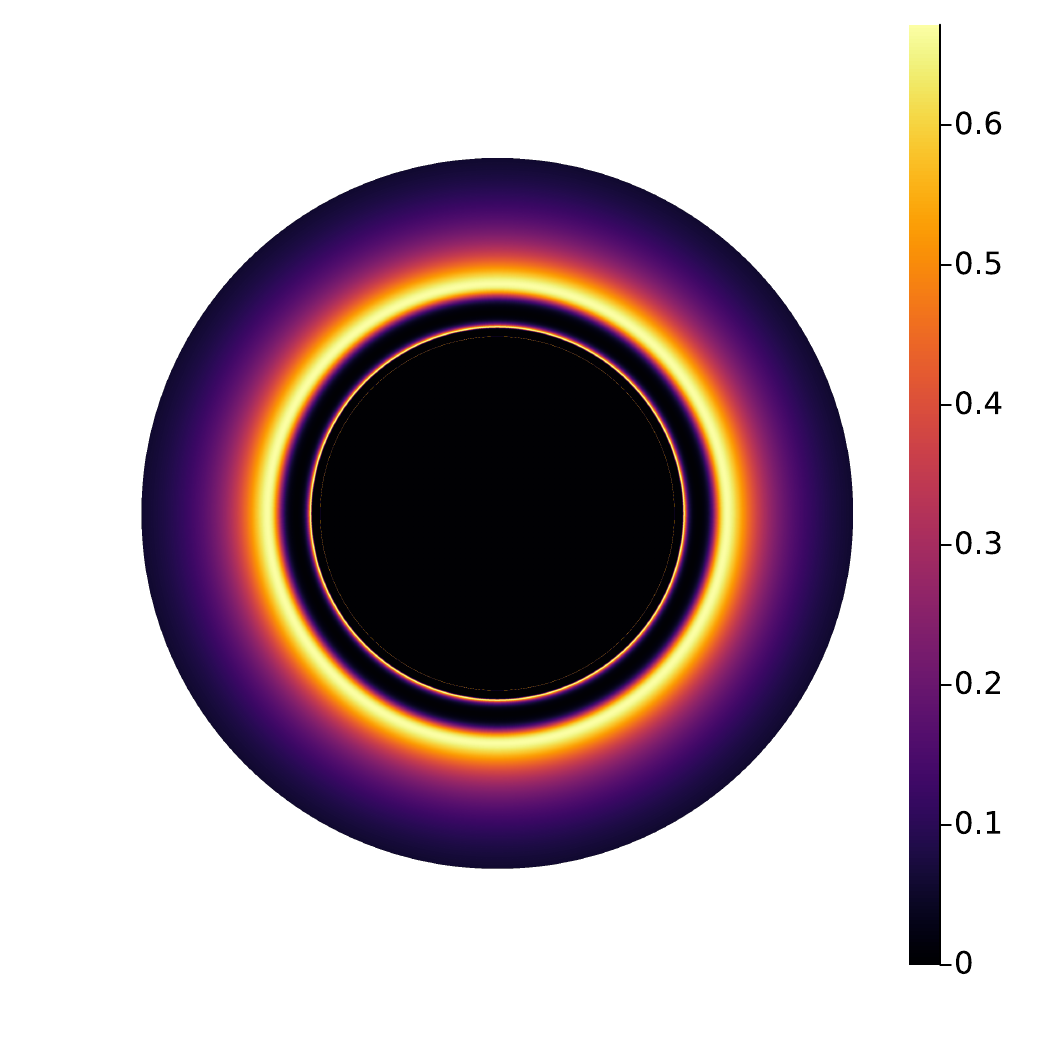}
    \includegraphics[width=0.32\linewidth]{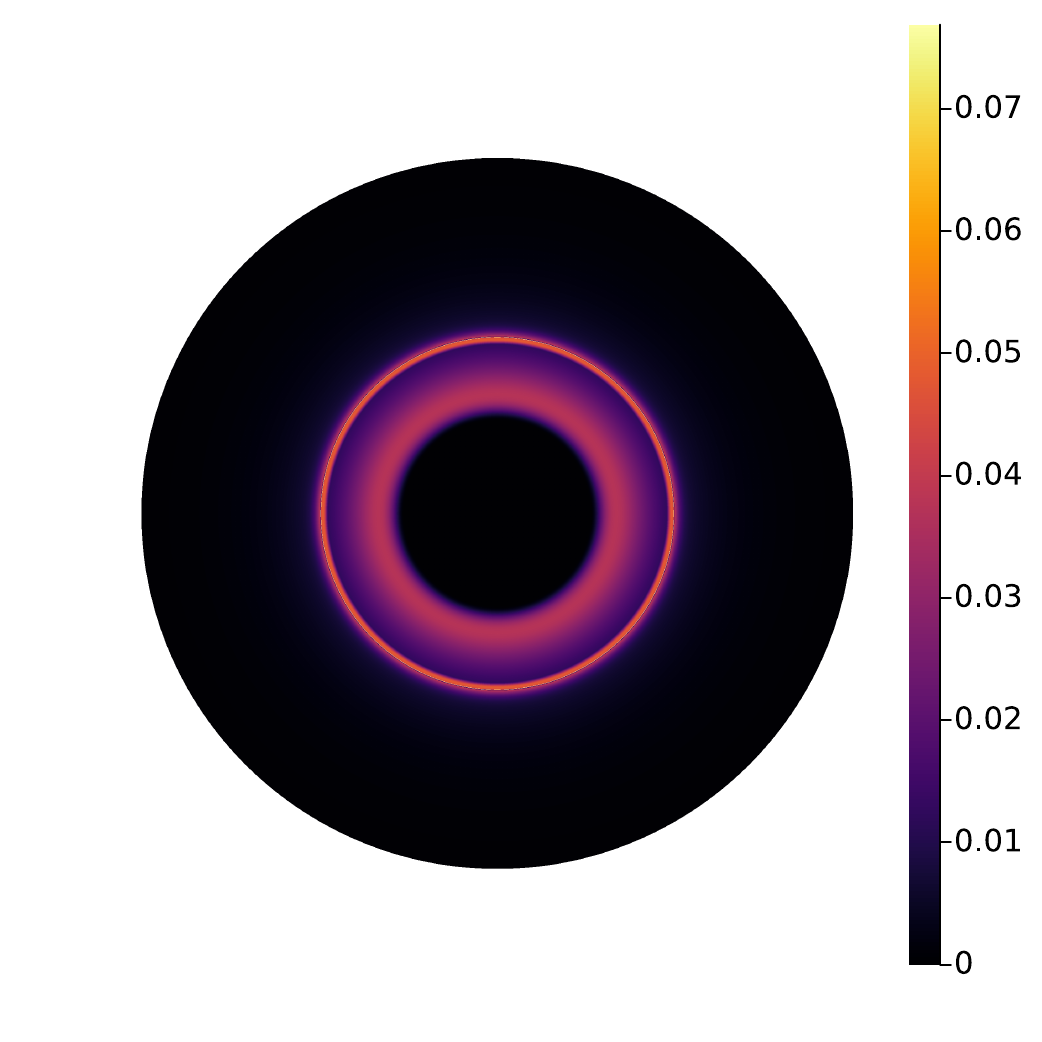}
    \includegraphics[width=0.32\linewidth]{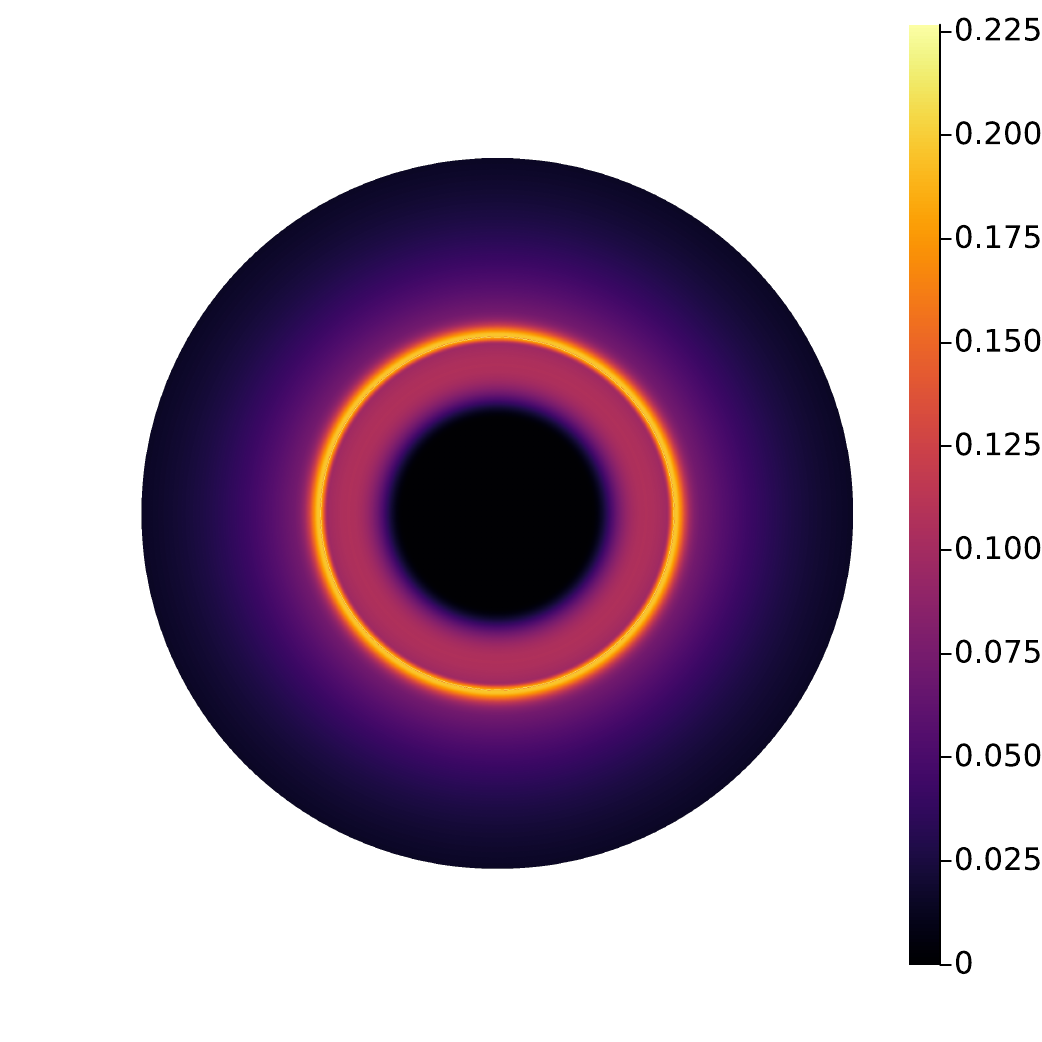}

    \includegraphics[width=0.32\linewidth]{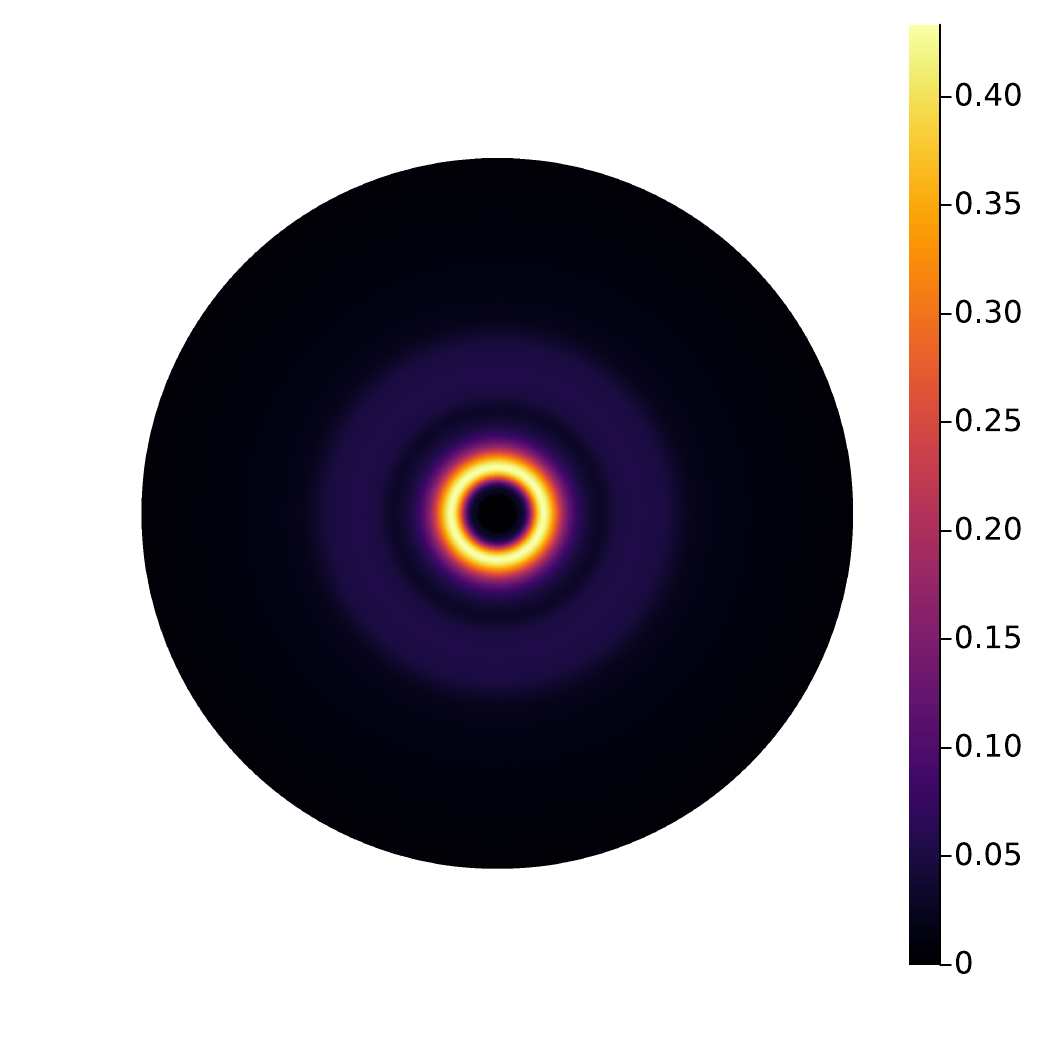}
    \includegraphics[width=0.32\linewidth]{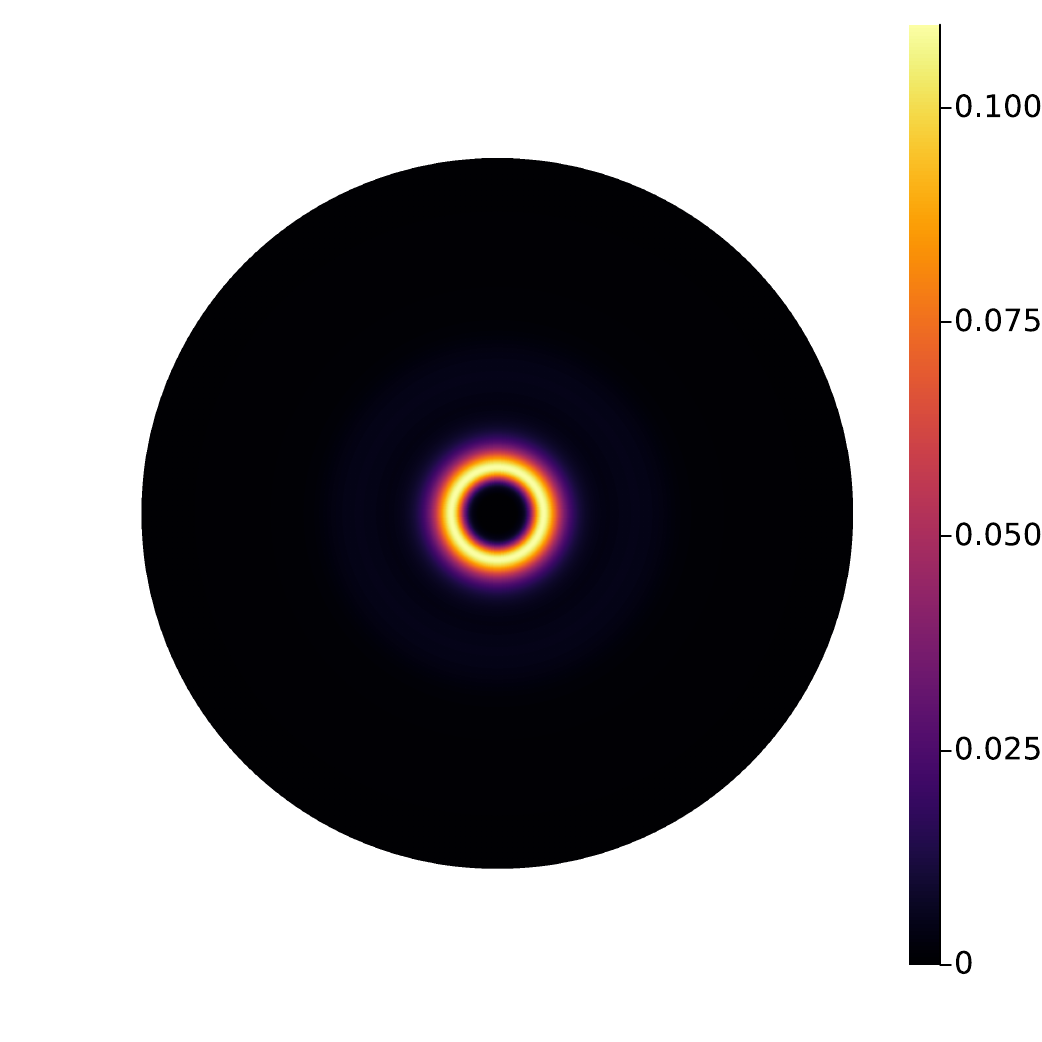}
    \includegraphics[width=0.32\linewidth]{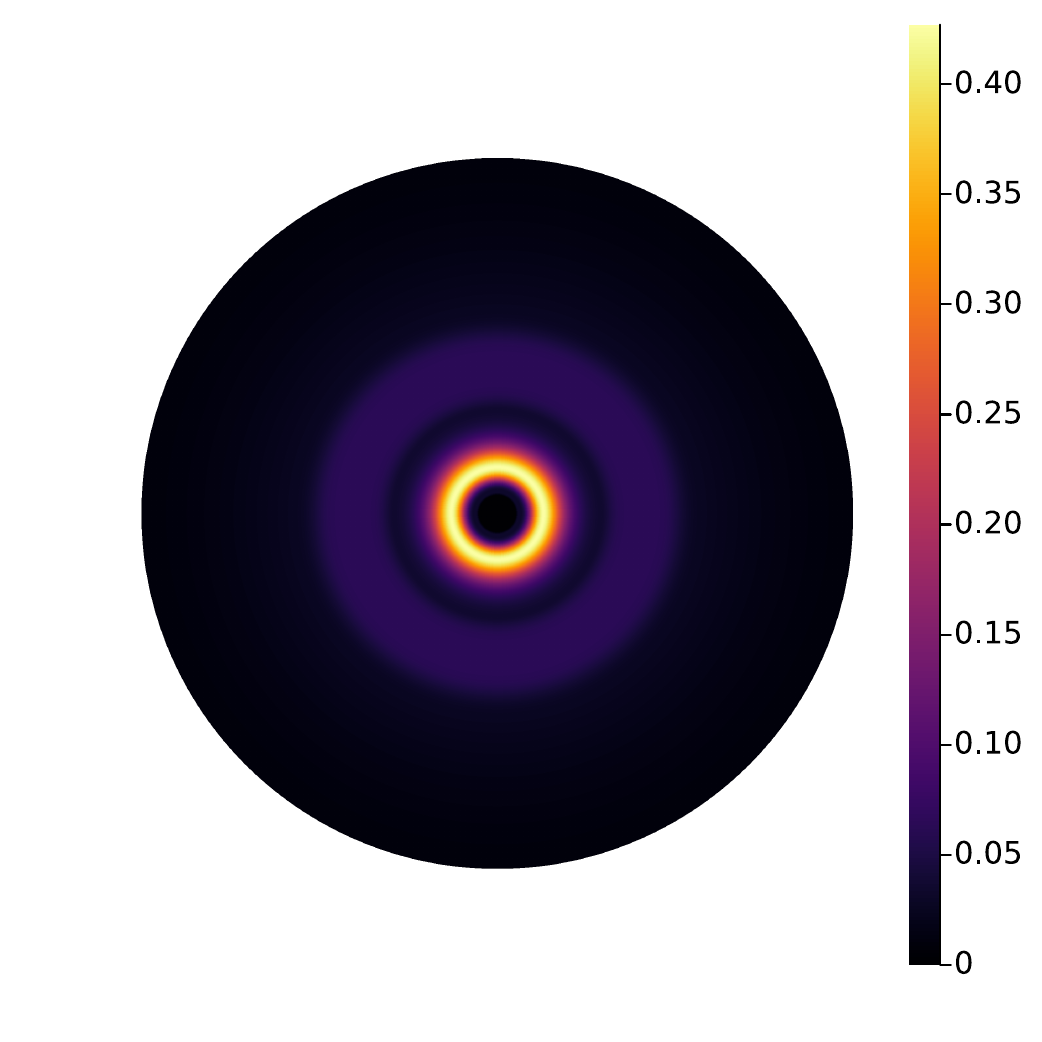}
    \caption{The top panels show the observational appearance of a black hole with parameters $M = 1$, $Q = 0.5$, $l = -5$ and $A = 6$. The bottom panels show the observational appearance of a naked singularity with parameters $M = 1$, $Q = 0.5$, $l = -10$ and $A = 3$.}
    \label{fig: images}
\end{figure}

Figure~\ref{fig: absortion} shows the observed intensity profiles $I_{\rm obs}(b)$ and Fig.~\ref{fig: images} the corresponding ray-traced images of the spacetime~\eqref{fmetric} for the three emission models. The images reveal how the Yukawa correction modifies the structure of photon trajectories and, consequently, the brightness distribution around the compact object.

\paragraph{Black hole case.}
For the black hole ($M=1$, $Q=0.5$, $l=-5$, $A=6$; top rows), the intensity profiles in Fig.~\ref{fig: absortion} vanish for $b<b_c$, which corresponds to the central dark region (the shadow) in Fig.~\ref{fig: images}. The profiles are dominated by a broad maximum produced by the direct emission ($m=1$), followed by a slowly decaying tail that follows the fall-off of $\tilde I(r)$ at large radii. The narrow spike near $b\simeq b_c$ is the signature of the lensed and photon-ring contributions ($m\geq 2$): these rays cross the disk close to the photon sphere, so the ring is thin in $b$, but it is amplified by the strong gravitational lensing and appears as a sharp feature in the profile and as a thin bright circle in the image. The relative strength of this feature depends strongly on the emission model. It is most prominent in panel (a), where the emissivity peaks close to the horizon and the direct emission is concentrated inside the ring; it is weaker in panel (b) and nearly merged with the direct emission in panel (c), where the emissivity is shifted to larger radii and the profile shows only a shoulder instead of an isolated spike. % CHECK against the GLM parameters
In the images, the bright annulus is the direct emission, and the thin inner ring is the lensing ring, in agreement with the standard picture for a spacetime with a single photon sphere~\cite{Gralla2019}.

\paragraph{Naked singularity case.}
The naked singularity ($M=1$, $Q=0.5$, $l=-10$, $A=3$; bottom rows) behaves qualitatively differently. Since there is no horizon, rays with small $b$ are not absorbed and can probe the region close to the center, where the emissivity is large. The profiles therefore show a pronounced peak at small $b$ together with a weak secondary bump at larger $b$, instead of a central gap followed by a sharp ring. Correspondingly, the images in Fig.~\ref{fig: images} show a compact, bright central core with only a small central depression (much smaller than the BH shadow) surrounded by a faint extended halo. The sharp photon-ring spike is absent, or strongly reduced, and the image is much more compact than in the black hole case. This gives a qualitative observational distinction between the two configurations, although it depends on the assumed emission model and on the behavior of the disk near the center.

\paragraph{Effect of the contribution parameters.}
The parameter $A$ controls the effective range of the Yukawa term and therefore regulates how strongly the Yukawa sector affects photon propagation. As the interaction range increases, departures from the RN image become progressively more pronounced: the critical curve, the position of the ring, and the relative brightness of the lensed features shift with respect to the RN case.

The ray-traced images visualize how the screened hair redistributes the lensed emission once an emissivity model is specified, complementing the quantitative shadow-radius constraint from the Event Horizon Telescope developed in Sec.~\ref{sec:eht} below.

\section{Quasinormal modes and ringdown signatures}\label{sec:qnm}

{
The optical analysis identifies the unstable photon region as the place where the finite-range sector is most efficiently probed. Ringdown provides a complementary question: not only where the critical null orbit is located, but how the geometry scatters and damps propagating fields around that region~\cite{Konoplya:2011qq,Konoplya:2016jvv}. This distinction is useful for the present solution because the Yukawa term is exponentially invisible at infinity. Any departure from the RN spectrum must therefore be generated in the strong-field region rather than by a change of the asymptotic mass.

{We restrict the discussion to massless scalar and electromagnetic test fields propagating on the fixed background geometry. This isolates a well-defined question: how much of the screened hair is actually seen by waves propagating on the background geometry?}

\subsection{Scattering problem and strong-field interpretation}

After the usual spherical-harmonic decomposition, both sectors reduce to
\begin{equation}
\frac{d^2\Psi}{dr_*^2}+\left[\omega^2-V_s(r)\right]\Psi=0,
\qquad
\frac{dr_*}{dr}=\frac{1}{f(r)},
\label{masterwave}
\end{equation}
with $s=0$ for a scalar field and $s=1$ for an electromagnetic test field. Their potentials can be written in the compact form
\begin{equation}
V_s(r)=f(r)\left[\frac{\ell(\ell+1)}{r^2}
+\left(1-s^2\right)\frac{f'(r)}{r}\right],
\qquad s=0,1,
\label{testpotentials}
\end{equation}
where $f(r)$ is given by Eq.~\eqref{fmetric}. Quasinormal boundary conditions select waves that are ingoing at the event horizon and outgoing at infinity. We write $\omega=\omega_R-i\omega_I$, with $\omega_I>0$ for a damped mode.

Equation~\eqref{testpotentials} contains the main physics more transparently than its fully expanded Yukawa form. The electromagnetic sector samples the Yukawa term directly through $f(r)$, whereas the scalar field also responds to its radial gradient. The two fields therefore interrogate the same screened geometry in slightly different ways at finite $\ell$. In both cases the potential vanishes at the event horizon and at spatial infinity and develops, for the black hole configurations considered below, a single dominant barrier outside $r_+$. The quasinormal spectrum is controlled primarily by the height, position and curvature of this barrier. The Yukawa sector can thus modify $\omega_R$ and $\omega_I$ differently: raising the barrier need not imply a proportionate change of its curvature.

This observation is especially important here because $A$ and $l$ play physically different roles. The amplitude $l$ fixes the unscreened strength of the additional sector, whereas $A^{-1}$ fixes the distance over which that strength survives. A large $|l|$ can therefore leave a surprisingly small spectral imprint if the exponential has already decayed before the potential reaches its maximum. Conversely, a moderate amplitude can be dynamically important when its range overlaps the photon-sphere region. Ringdown is consequently sensitive not to $l$ in isolation, but to the \emph{residual hair at the scattering barrier}.

{The frequencies were evaluated with the semianalytic WKB expansion around the maximum of the corresponding potential, following the foundational third-order construction of Schutz, Will and Iyer~\cite{Schutz:1985km,Iyer:1986np,Iyer:1987ny} and extended to sixth order following Konoplya~\cite{Konoplya:2003ii}, consistently with the general framework reviewed in Refs.~\cite{Konoplya:2011qq,Matyjasek:2017psv,Konoplya:2019hlu}. This approach has become a standard tool for black holes in higher dimensions, in higher-curvature theories and for regular geometries~\cite{Konoplya:2003dd,Konoplya:2005et,Konoplya:2008ix,Konoplya:2020bxa,Konoplya:2020juj,Konoplya:2023ahd}, and its connection with grey-body factors has recently been clarified~\cite{Konoplya:2024lir}. Tables~\ref{tab:qnm_scalar_main}--\ref{tab:qnm_overtones} report the sixth-order value; for the fundamental mode we also show the third-order result, which agrees with it to better than $0.2\%$ in $\mathrm{Re}(M\omega)$ and to a few percent in $\mathrm{Im}(M\omega)$ throughout the parameter range studied.

To validate the WKB frequencies independently, and in particular the overtones, we also computed the spectrum with a pseudospectral collocation method on hyperboloidal slices~\cite{Ansorg:2016ztf,Jansen:2017oag}. In this formulation the quasinormal boundary conditions are enforced by the regularity of the eigenfunctions at the event horizon and at future null infinity, no tortoise coordinate is needed, and the frequencies are obtained as the eigenvalues of a quadratic eigenproblem. The two methods agree remarkably well: for $\ell=2$ the fundamental scalar and electromagnetic modes coincide to better than $0.18\%$ in $\mathrm{Re}(M\omega)$ and $0.21\%$ in $\mathrm{Im}(M\omega)$ for all the configurations of Table~\ref{tab:qnm_scalar_main}, and the first overtone to $0.62\%$ and $0.37\%$. The comparison for $\ell=3$ and $n=0,1,2$ is reported in Table~\ref{tab:qnm_overtones}. The agreement certifies that the reported frequencies, including the overtones, are reliable for the Yukawa-type geometry. \textcolor{blue}{The formulation, the discretization and the convergence tests are given in Appendix~\ref{app:spectral}.}

For all parameter sets used in the numerical analysis the exterior potentials remain non-negative, which excludes growing modes in the scalar and electromagnetic test-field problems on the fixed background.}

The connection with the optical sector becomes particularly transparent at large $\ell$. The peak of both potentials approaches the unstable null orbit and the spectrum obeys
\begin{equation}
\omega_{\ell n}\simeq
\ell\,\Omega_c-i\left(n+\frac12\right)\lambda,
\qquad
\Omega_c=\frac{\sqrt{f(r_{\rm ph})}}{r_{\rm ph}},
\label{eikonal}
\end{equation}
where $\lambda$ is the Lyapunov exponent of the circular null orbit. Since $R_{\rm sh}=\Omega_c^{-1}$, the shadow and the real part of the eikonal spectrum are two manifestations of the same local geometry; this correspondence is not universal in every gravity theory~\cite{Konoplya:2017wot}, but it holds for the Einstein--Maxwell-matter setting considered here, as the finite-$\ell$ results below confirm. This correspondence will be useful below, but the finite-$\ell$ results contain additional information because the scalar potential, in particular, samples derivatives of $f(r)$ away from the exact photon orbit.

\subsection{Numerical spectrum: what the ringdown actually measures}

We take $M=1$ and $Q=0.5$ as a representative charged background and use RN, obtained at $l=0$, as the reference geometry. Table~\ref{tab:qnm_scalar_main} shows the fundamental scalar mode with $\ell=2$. The first message from the table is that the spectral response is not monotonic in the bare magnitude of the hair. At $A=1$ and $l=-1$, for example, $\mathrm{Re}(M\omega_{20})$ rises from the RN value $0.506$ to approximately $0.623$, a shift of about $23\%$, while the damping changes only mildly. The geometry has therefore altered the characteristic oscillation scale much more strongly than the decay time. By contrast, the case $A=2$, $l=-2$ gives $M\omega_{20}\simeq0.563-0.074i$: the real frequency is again displaced, but now the most conspicuous effect is the smaller magnitude of the imaginary part and hence a longer-lived test-field response.

\begin{table*}[t]
\centering
\caption{{Fundamental scalar quasinormal frequency for $M=1$, $Q=0.5$, $n=0$ and $\ell=2$. Here $r_{\rm max}$ is the maximum of the scalar scattering barrier and $M\omega_{20}$ is the sixth-order WKB value~\cite{Konoplya:2003ii}; $M\omega_{20}^{(3)}$ is the third-order value~\cite{Schutz:1985km,Iyer:1986np} shown alongside it as a convergence check.}}
\label{tab:qnm_scalar_main}
\begin{ruledtabular}
\begin{tabular}{ccccccc}
$A$ & $l$ & $r_+$ & $r_{\rm max}$ & $M\omega_{20}^{(3)}$ & $M\omega_{20}$ & $R_{\rm sh}/M$\\
\hline
RN & 0 & 1.866025 & 2.781630 & {$0.506257-0.101533i$} & {$0.505528-0.097835i$} & 4.967914 \\
0.5 & -0.5 & 1.487141 & 2.247721 & {$0.598315-0.111175i$} & {$0.597651-0.107541i$} & 4.198380 \\
1.0 & -0.5 & 1.624094 & 2.465867 & {$0.548354-0.101082i$} & {$0.547788-0.097968i$} & 4.584997 \\
2.0 & -0.5 & 1.774298 & 2.688592 & {$0.515677-0.098310i$} & {$0.515089-0.095182i$} & 4.880489 \\
4.0 & -0.5 & 1.852774 & 2.774690 & {$0.506759-0.100868i$} & {$0.506037-0.097178i$} & 4.963047 \\
1.0 & -1.0 & 1.275531 & 2.010990 & {$0.623267-0.098814i$} & {$0.622889-0.096404i$} & 4.028111 \\
2.0 & -1.0 & 1.658969 & 2.575213 & {$0.527215-0.093999i$} & {$0.526770-0.091471i$} & 4.776328 \\
4.0 & -1.0 & 1.838752 & 2.767552 & {$0.507262-0.100180i$} & {$0.506546-0.096489i$} & 4.958067 \\
2.0 & -2.0 & 1.146220 & 2.202222 & {$0.563200-0.075652i$} & {$0.562978-0.073977i$} & 4.460192 \\
4.0 & -2.0 & 1.807933 & 2.752629 & {$0.508268-0.098730i$} & {$0.507558-0.095007i$} & 4.947745 \\
4.0 & -4.0 & 1.729658 & 2.719720 & {$0.510225-0.095478i$} & {$0.509505-0.091551i$} & 4.925423 \\
\end{tabular}
\end{ruledtabular}
\end{table*}

{We note that the strong-screening benchmark $A=4$ used throughout this section corresponds to $Ar_+\gtrsim7$, outside the radial domain $Ar\le4$ in which the dominant energy condition holds (Sec.~\ref{sec:yukawa}); the WEC, NEC and SEC, on which the present analysis relies, remain globally satisfied there.}

The second, and more characteristic, result is the rapid spectral return to RN when the range is shortened. Keeping $l=-1$ fixed while increasing $A$ from $1$ to $4$ drives the real part from approximately $0.623$ to $0.507$, essentially restoring the RN value despite the fact that the amplitude $l$ has not been reduced. Figure~\ref{fig:qnm_scan} shows the same behavior over the parameter scan. The different $A$ curves separate when the Yukawa range overlaps the strong-field region and collapse toward RN when the Yukawa term is screened before reaching the barrier. The imaginary part exhibits a richer response because it is sensitive to the curvature of the potential rather than only to its height.

\begin{figure*}[t]
\centering
\includegraphics[width=0.47\textwidth]{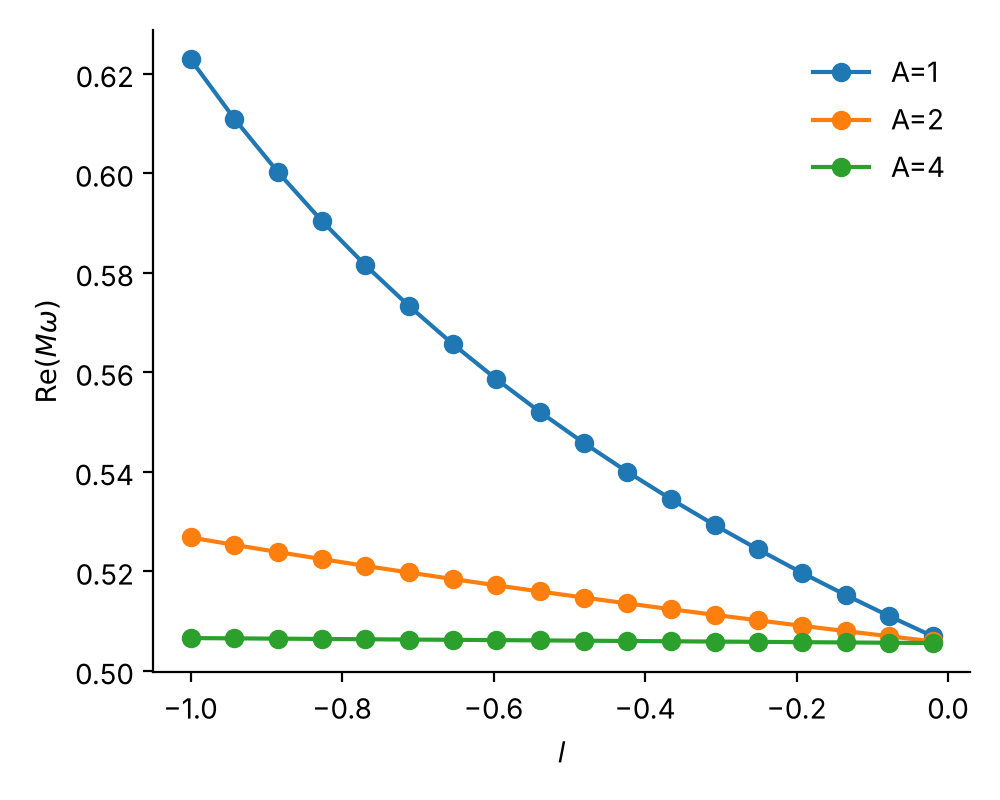}
\hfill
\includegraphics[width=0.47\textwidth]{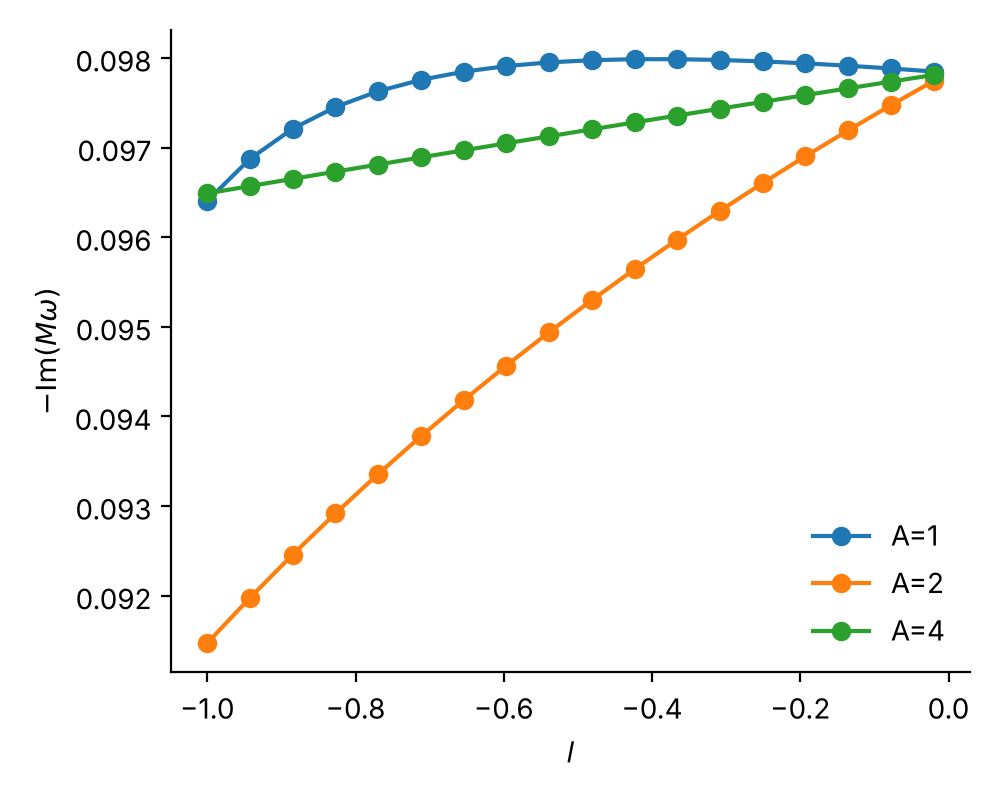}
\caption{{Fundamental scalar mode with $n=0$ and $\ell=2$ for $M=1$ and $Q=0.5$, sixth-order WKB~\cite{Konoplya:2003ii}. Left: oscillation frequency. Right: damping rate. The approach of the different curves to the RN values as $A$ increases illustrates that the relevant quantity is the amount of Yukawa hair that survives in the scattering region, rather than the bare amplitude $l$ alone.}}
\label{fig:qnm_scan}
\end{figure*}

The separate plots of the oscillation and damping rates are complemented by the
complex-frequency plane. Figure~\ref{fig:qnm_complex} shows that varying the
screening parameters does not move the modes along a universal one-dimensional
trajectory. Some configurations are displaced mainly in the real direction,
whereas others show a substantial change in the imaginary part. This is the
spectral counterpart of the fact that the Yukawa sector changes not only the
height of the barrier but also its curvature and width. The return toward the RN
point as screening becomes stronger is therefore a genuinely complex spectral
flow rather than a simple rescaling of the frequency.

\begin{figure}[t]
\centering
\includegraphics[width=0.58\textwidth]{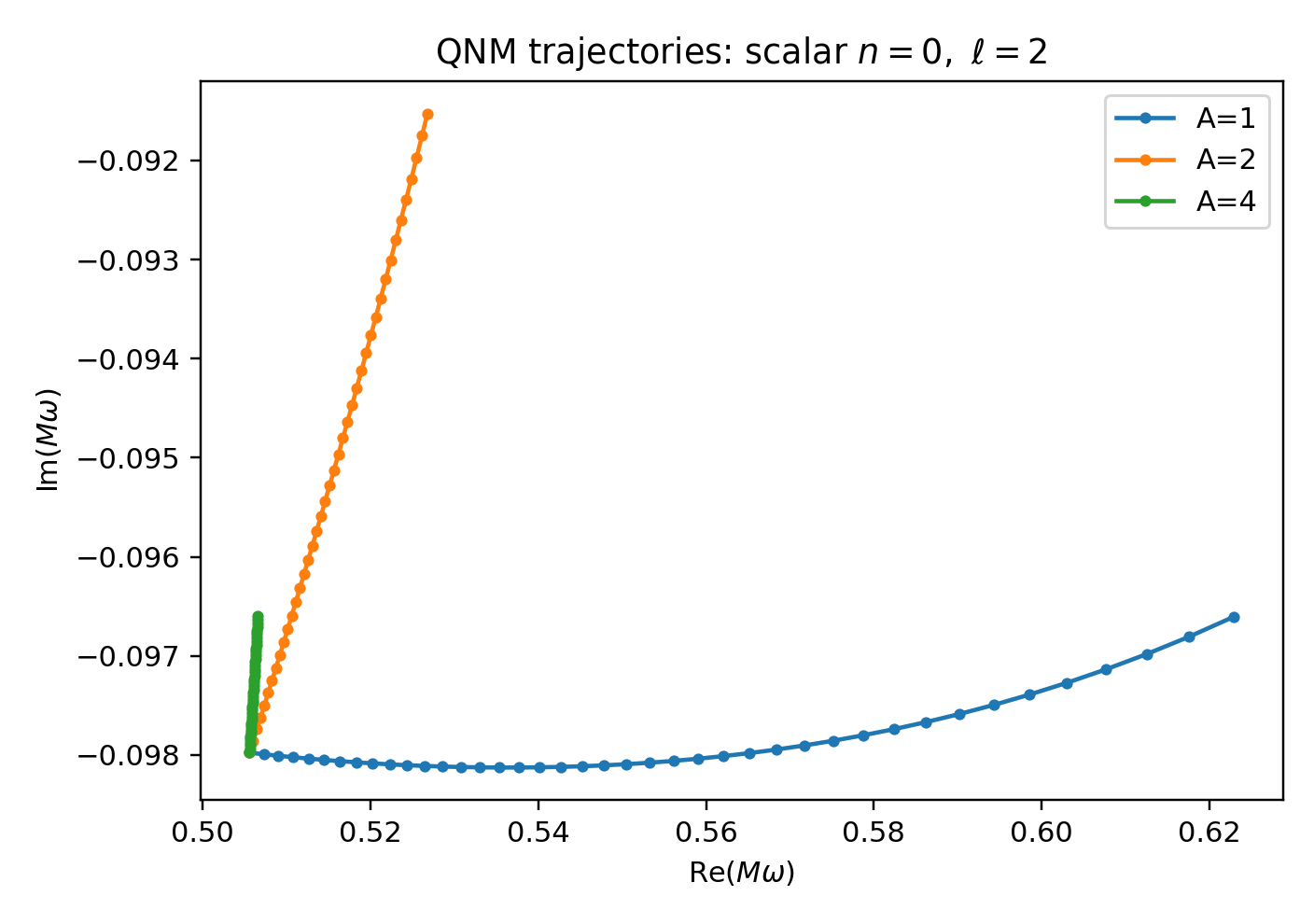}
\caption{{Fundamental scalar $\ell=2$ quasinormal frequencies
in the complex plane for representative Yukawa configurations with $M=1$ and
$Q=0.5$. The different directions of the spectral displacement show that the
finite-range contribution affects the oscillation and damping scales differently,
while strong screening drives the modes back toward the RN reference point.}}
\label{fig:qnm_complex}
\end{figure}

The optical analysis already identified the residual hair
$\chi_{\rm ph}$ in Eq.~\eqref{effectivehair} as a local measure of how much of
the Yukawa correction survives at the photon sphere. The same quantity is
naturally relevant to the scattering barrier. When $A r_{\rm ph}\gg1$, $\chi_{\rm ph}$ is exponentially small and the geometry sampled by the barrier becomes RN-like even when $|l|$ itself is not. This explains the apparent degeneracy visible in the numerical spectrum and gives a more physical parametrization of the ringdown phenomenology than the bare pair $(A,l)$ considered separately.

The potential profiles make the same point from the scattering perspective. As shown in Fig.~\ref{fig:qnm_potential}, long-range hair moves the maximum and changes both its height and width, whereas strong screening almost restores the RN barrier. The WKB response then has an immediate interpretation: changes of the height predominantly feed the oscillation scale, while changes of the curvature control the leakage time through the barrier. This is why configurations with comparable shifts in $\omega_R$ need not possess comparable damping rates.

\begin{figure}[t]
\centering
\includegraphics[width=0.58\textwidth]{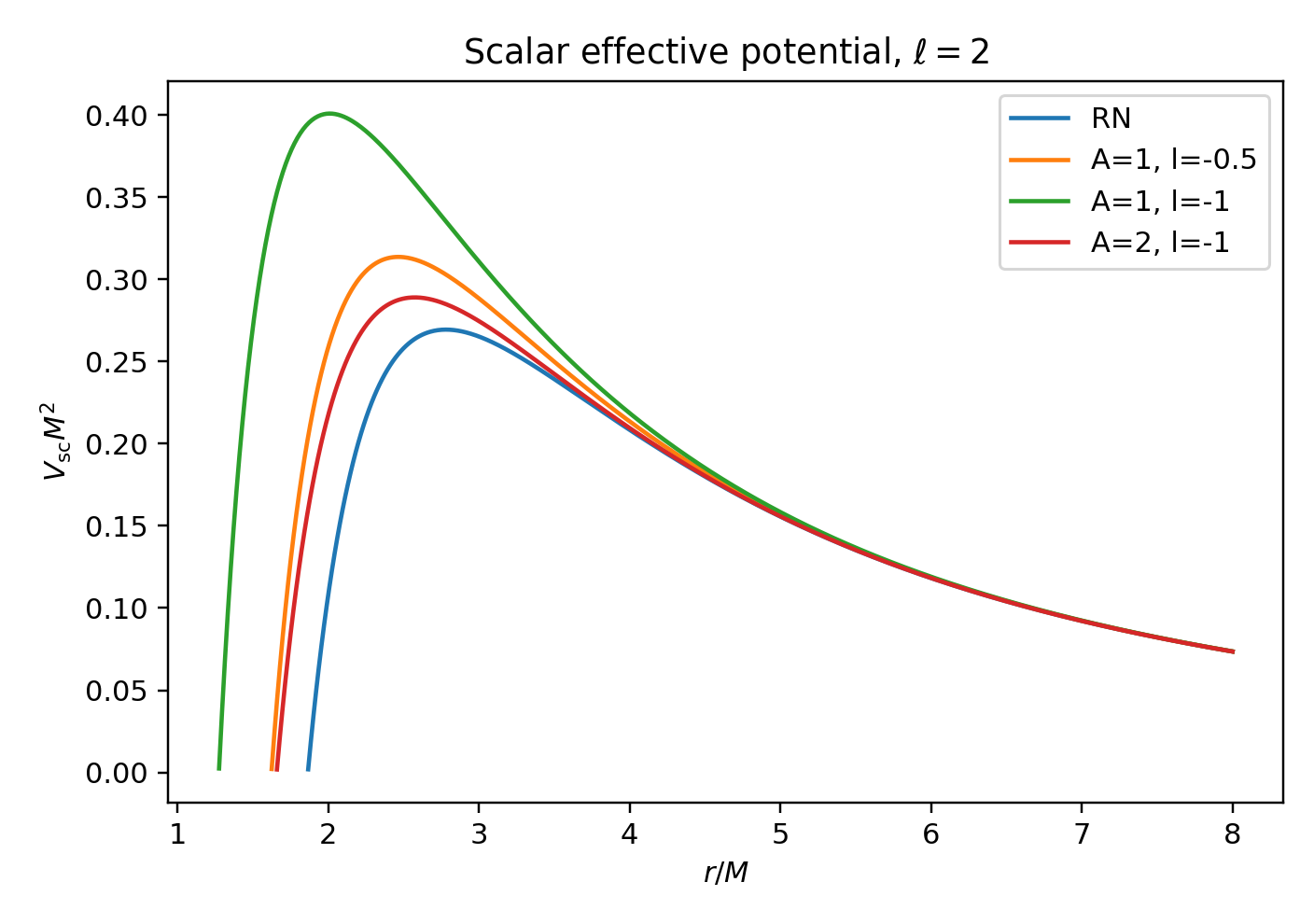}
\caption{{Scalar effective potential for $\ell=2$, $M=1$ and $Q=0.5$. The Yukawa sector changes the location, height and curvature of the scattering barrier when its range overlaps the strong-field region; sufficiently strong screening restores the RN profile.}}
\label{fig:qnm_potential}
\end{figure}

The electromagnetic sector deserves a separate numerical check rather than being
treated only as a formal companion to the scalar field.  Because
$V_{\rm em}=f\ell(\ell+1)/r^2$, it removes the explicit $f'(r)/r$ term and
therefore provides a cleaner probe of the metric function itself.  Figure~\ref{fig:qnm_vector_scan}
shows that the same screening mechanism survives this change of probe.  For
small and moderate $A$, changing $l$ produces visible shifts of both the
oscillation and damping scales, whereas for large $A$ the curves are driven
back toward the RN reference values.  The recovery of RN is therefore not a
peculiarity of the scalar potential.

\begin{figure*}[t]
\centering
\includegraphics[width=0.47\textwidth]{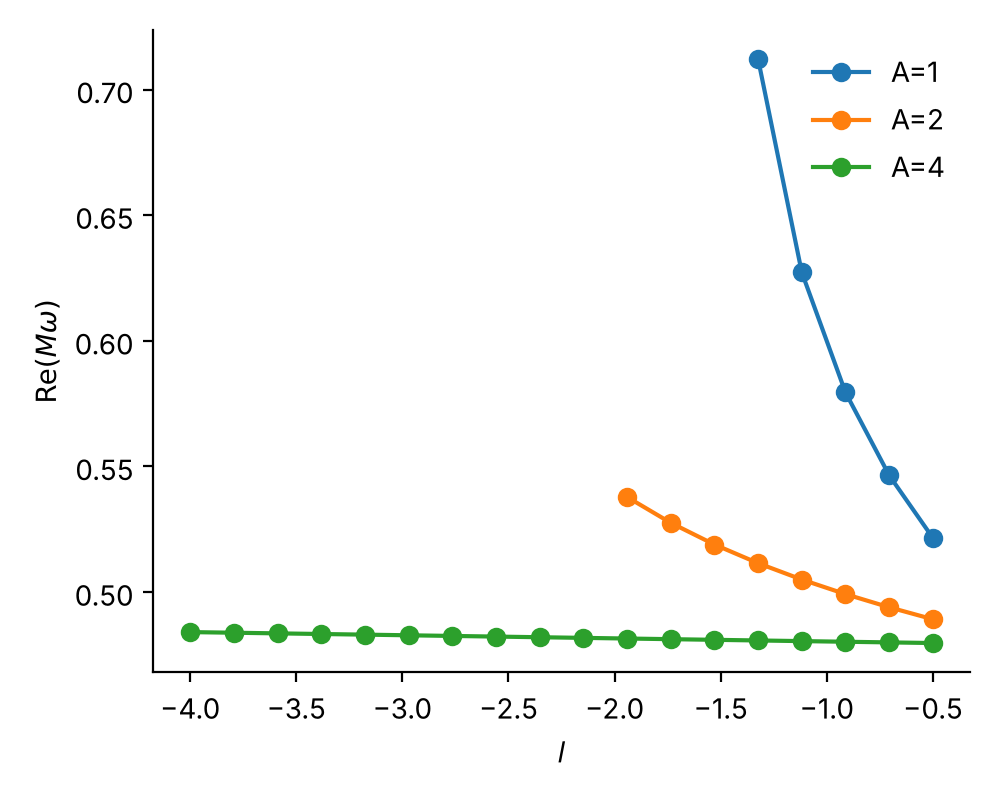}
\hfill
\includegraphics[width=0.47\textwidth]{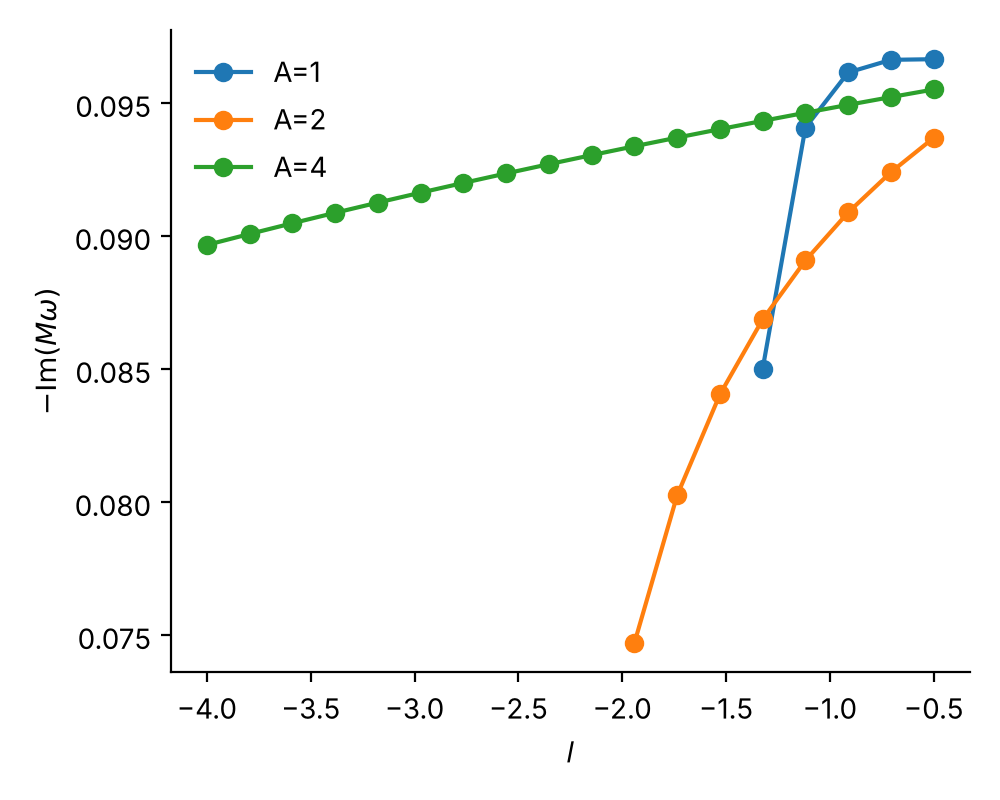}
\caption{{Fundamental electromagnetic test-field mode with
$n=0$ and $\ell=2$ for $M=1$ and $Q=0.5$, sixth-order WKB~\cite{Konoplya:2003ii}. Left: oscillation frequency.
Right: damping rate. The vector sector displays the same finite-range screening
seen for the scalar field: when the Yukawa contribution decays before the
scattering region is reached, the spectrum approaches its RN value.}}
\label{fig:qnm_vector_scan}
\end{figure*}

A direct scalar--vector comparison is more informative than quoting the two
spectra independently.  Table~\ref{tab:qnm_scalar_vector} displays the
fundamental modes over several multipoles.  At low $\ell$ the scalar and vector
frequencies differ because the scalar potential contains the additional
$f'(r)/r$ term.  The difference becomes progressively less important compared
with the overall oscillation scale as $\ell$ grows.  This is precisely what is
expected if both sectors are converging toward a common photon-sphere-controlled
eikonal regime.

\begin{table*}[t]
\centering
\caption{{Comparison of the fundamental scalar and
electromagnetic test-field frequencies for representative geometries.  We set
$M=1$, $Q=0.5$ and $n=0$.  The entries are sixth-order WKB values~\cite{Konoplya:2003ii}.}}
\label{tab:qnm_scalar_vector}
\begin{ruledtabular}
\begin{tabular}{cccc}
Geometry & $\ell$ & $M\omega_{\rm sc}$ & $M\omega_{\rm em}$\\
\hline
RN & 1 & {$0.3048-0.0987i$} & {$0.2590-0.0940i$} \\
 & 2 & {$0.5055-0.0978i$} & {$0.4791-0.0962i$} \\
 & 3 & {$0.7063-0.0976i$} & {$0.6876-0.0968i$} \\
 & 4 & {$0.9072-0.0976i$} & {$0.8927-0.0971i$} \\
 & 5 & {$1.1083-0.0975i$} & {$1.0964-0.0972i$} \\
$A=1,\ l=-0.5$ & 1 & {$0.3306-0.0985i$} & {$0.2853-0.0948i$} \\
 & 2 & {$0.5478-0.0980i$} & {$0.5216-0.0967i$} \\
 & 3 & {$0.7653-0.0978i$} & {$0.7467-0.0972i$} \\
 & 4 & {$0.9830-0.0978i$} & {$0.9686-0.0974i$} \\
 & 5 & {$1.2008-0.0978i$} & {$1.1891-0.0975i$} \\
$A=1,\ l=-1$ & 1 & {$0.3755-0.0960i$} & {$0.3319-0.0940i$} \\
 & 2 & {$0.6229-0.0964i$} & {$0.5975-0.0956i$} \\
 & 3 & {$0.8706-0.0965i$} & {$0.8526-0.0960i$} \\
 & 4 & {$1.1185-0.0965i$} & {$1.1045-0.0962i$} \\
 & 5 & {$1.3665-0.0965i$} & {$1.3551-0.0963i$} \\
$A=4,\ l=-1$ & 1 & {$0.3053-0.0973i$} & {$0.2600-0.0925i$} \\
 & 2 & {$0.5065-0.0965i$} & {$0.4804-0.0948i$} \\
 & 3 & {$0.7077-0.0963i$} & {$0.6892-0.0955i$} \\
 & 4 & {$0.9091-0.0962i$} & {$0.8947-0.0957i$} \\
 & 5 & {$1.1105-0.0962i$} & {$1.0988-0.0959i$} \\
\end{tabular}
\end{ruledtabular}
\end{table*}

This convergence can be displayed without relying only on the raw frequencies.
Figure~\ref{fig:qnm_eikonal_compare} shows the scaled real part
$\mathrm{Re}(M\omega)/(\ell+1/2)$.  Scalar and electromagnetic sequences move
toward the same geometry-dependent scale as $\ell$ increases.  The result gives
a useful internal check: the finite-$\ell$ field dependence is real, but the
dominant high-multipole response is controlled by the background null orbit.

\begin{figure}[t]
\centering
\includegraphics[width=0.62\textwidth]{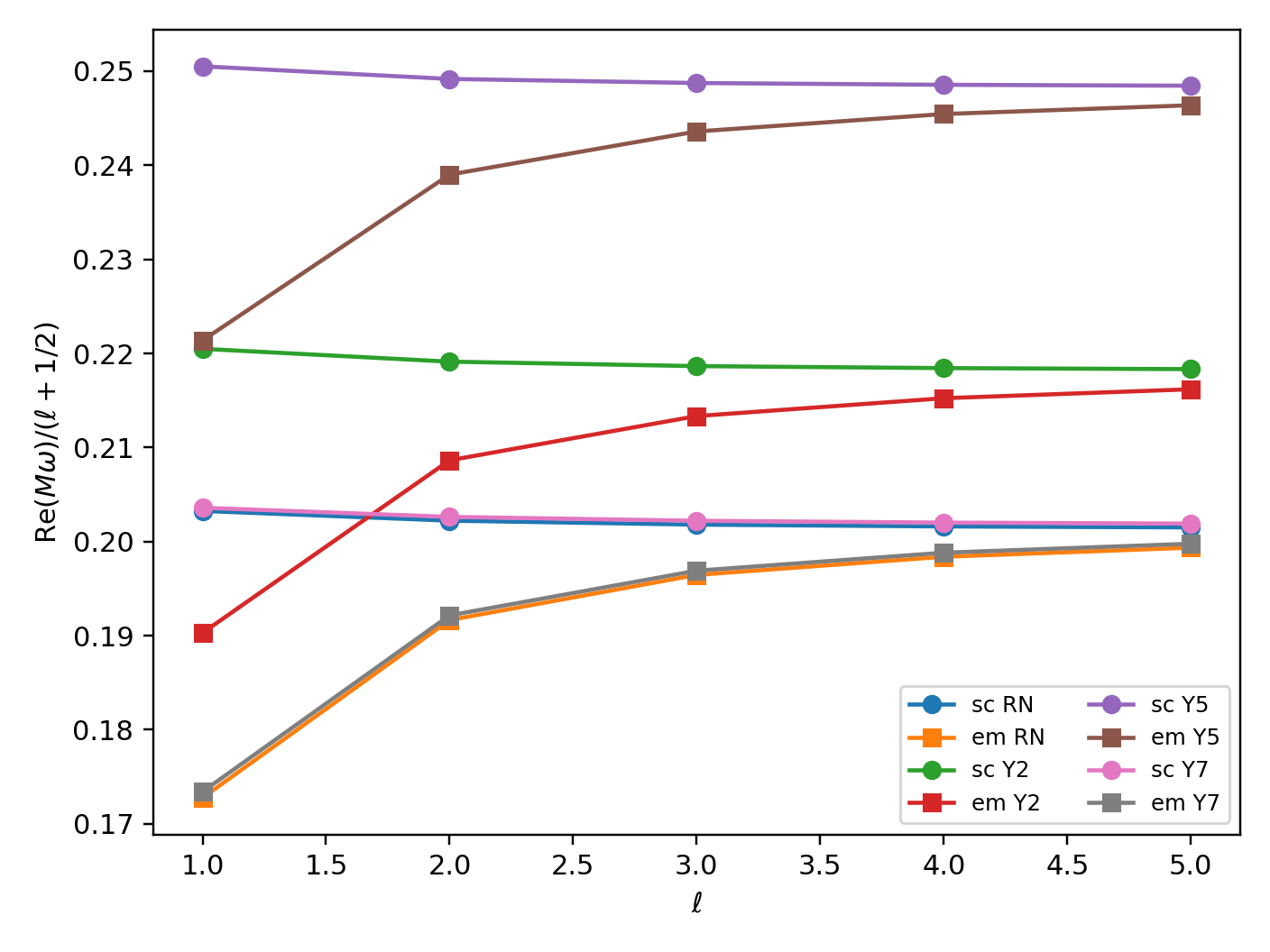}
\caption{{Multipole dependence of the scaled real frequency
for scalar and electromagnetic fundamental modes in representative geometries.
The decreasing separation between the two test fields illustrates their
approach to the common photon-sphere-controlled eikonal behavior.}}
\label{fig:qnm_eikonal_compare}
\end{figure}

The fundamental mode is the cleanest quantity for comparing geometries, but the
first overtones test whether the same conclusion persists deeper in the
ringdown spectrum.  Table~\ref{tab:qnm_overtones} gives the $\ell=3$ sequence
for two representative Yukawa configurations, comparing the sixth-order WKB values with the spectral ones.  Increasing $n$ produces the
expected stronger damping, while the contribution remains visible in both
sectors; the two methods agree to better than $0.03\%$ ($0.09\%$) in the real (imaginary) part for $n=0$, $0.08\%$ ($0.18\%$) for $n=1$ and $0.11\%$ ($0.62\%$) for $n=2$. The table shows that the screened
hair is not encoded exclusively in the fundamental mode.

\begin{table*}[t]
\centering
\caption{First scalar and electromagnetic overtones for $\ell=3$, $M=1$ and $Q=0.5$. For each field the sixth-order WKB value~\cite{Konoplya:2003ii} is compared with the result of an independent pseudospectral calculation on hyperboloidal slices~\cite{Ansorg:2016ztf,Jansen:2017oag}, illustrating how the Yukawa contribution propagates through the overtone sequence.}
\label{tab:qnm_overtones}
\begin{ruledtabular}
\begin{tabular}{cccccc}
 & & \multicolumn{2}{c}{Scalar} & \multicolumn{2}{c}{Electromagnetic}\\
Geometry & $n$ & $M\omega_{\rm WKB6}$ & $M\omega_{\rm spec}$ & $M\omega_{\rm WKB6}$ & $M\omega_{\rm spec}$\\
\hline
RN & 0 & $0.7063-0.0976i$ & $0.7065-0.0977i$ & $0.6876-0.0968i$ & $0.6878-0.0969i$ \\
 & 1 & $0.6922-0.2953i$ & $0.6926-0.2957i$ & $0.6730-0.2929i$ & $0.6735-0.2933i$ \\
 & 2 & $0.6672-0.4982i$ & $0.6671-0.5013i$ & $0.6473-0.4944i$ & $0.6469-0.4972i$ \\
$A=1,\ l=-1$ & 0 & $0.8706-0.0965i$ & $0.8707-0.0965i$ & $0.8526-0.0960i$ & $0.8527-0.0961i$ \\
 & 1 & $0.8613-0.2905i$ & $0.8617-0.2910i$ & $0.8430-0.2893i$ & $0.8435-0.2898i$ \\
 & 2 & $0.8440-0.4871i$ & $0.8449-0.4893i$ & $0.8254-0.4854i$ & $0.8262-0.4875i$ \\
$A=4,\ l=-1$ & 0 & $0.7077-0.0963i$ & $0.7079-0.0964i$ & $0.6892-0.0955i$ & $0.6894-0.0955i$ \\
 & 1 & $0.6940-0.2913i$ & $0.6946-0.2916i$ & $0.6751-0.2888i$ & $0.6756-0.2891i$ \\
 & 2 & $0.6699-0.4914i$ & $0.6699-0.4938i$ & $0.6503-0.4875i$ & $0.6503-0.4901i$ \\
\end{tabular}
\end{ruledtabular}
\end{table*}

Taken together, the scalar and vector results separate two effects that
would otherwise be easy to conflate.  The detailed finite-$\ell$ spectrum retains
information about the spin-dependent shape of the scattering potential, while
the systematic return to RN under strong screening and the large-$\ell$
convergence are properties of the background geometry itself.

The same geometric origin links the ringdown shift to the shadow. Combining $R_{\rm sh}=\Omega_c^{-1}$ with Eq.~\eqref{eikonal} gives $\mathrm{Re}(\omega_{\ell n})R_{\rm sh}\simeq\ell$ at large $\ell$. The finite-$\ell$ scan already displays the corresponding anticorrelation: configurations that enlarge the characteristic oscillation frequency tend to reduce the shadow radius. Thus the optical and spectral observables should not be regarded as two unrelated signatures of the Yukawa sector. They are two projections of the same screened contribution around the photon sphere, with the damping rate supplying the additional information carried by the curvature of the scattering barrier.

Figure~\ref{fig:shadow_qnm} makes this common origin visible without invoking
the strict eikonal limit. Already for the fundamental $\ell=2$ scalar mode, the
fractional displacement of the real frequency tracks the change of the shadow
radius with the opposite sign expected from $R_{\rm sh}=\Omega_c^{-1}$. The
relation is not exact at finite $\ell$, nor should it be: the wave potential
samples a finite radial neighborhood of its maximum rather than only the null
orbit. Precisely this departure is useful, because it allows the ringdown spectrum
to retain information beyond the critical impact parameter.

\begin{figure}[t]
\centering
\includegraphics[width=0.58\textwidth]{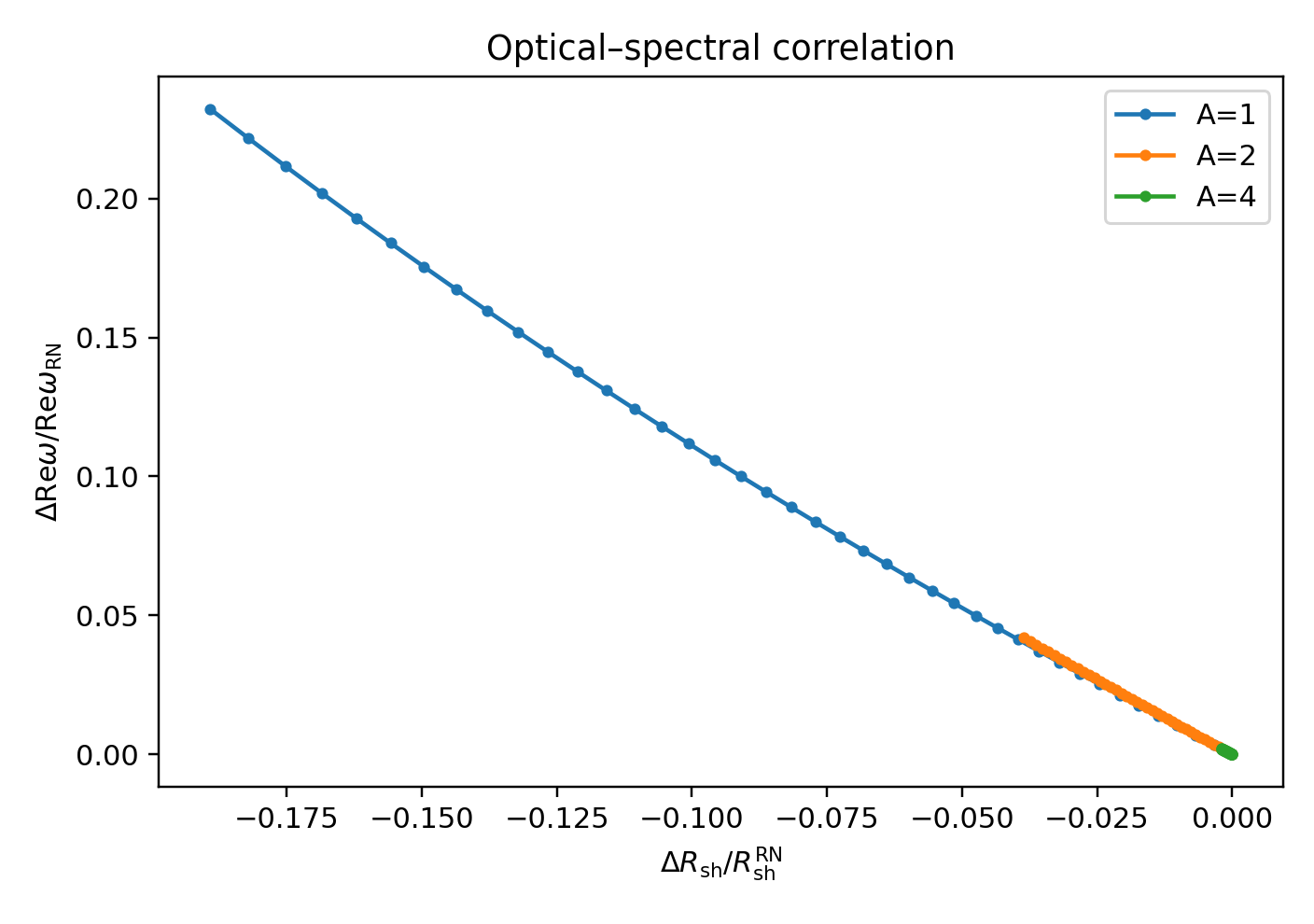}
\caption{{Correlation between the fractional change of the
shadow radius and that of the real part of the fundamental scalar $\ell=2$
frequency relative to RN. The anticorrelation is the finite-multipole precursor
of the eikonal relation $\mathrm{Re}(\omega_{\ell n})R_{\rm sh}\simeq\ell$ and
shows that the optical and spectral shifts originate in the same screened
strong-field geometry.}}
\label{fig:shadow_qnm}
\end{figure}

The physical picture that emerges is therefore simple. The parameters $(A,l)$ specify the matter sector globally, but ringdown only measures the part of that sector that reaches the strong-field scattering region. This produces a natural screening degeneracy: rather different microscopic pairs $(A,l)$ can yield nearly RN-like spectra if $|\chi_{\rm ph}|$ is small. Conversely, a long-range contribution can remain spectroscopically visible even for moderate $|l|$. Breaking this degeneracy requires combining observables that probe different aspects of the same region---for example the shadow radius, which is tied mainly to the critical impact parameter, with the complex QNM frequency, whose imaginary part also resolves the local instability timescale. In this sense the most interesting prediction of the finite-range sector is not simply a shifted QNM frequency, but a correlated optical--ringdown pattern controlled by how efficiently the Yukawa hair penetrates the photon-sphere region.
}

\section{Conclusion}\label{section8}

{
We have shown that a nonlocal relation between the radial pressure and the matter enclosed within each radius, supplemented only by the requirement that the geometry be of Kerr--Schild type, fixes the form of a charged black hole within ordinary Einstein--Maxwell gravity. The additional sector is exponentially screened, and the Yukawa profile is the outcome of the closure itself rather than an input. The solution is exact and sourced by an effective anisotropic fluid. It shows that a finite-range correction to the RN geometry can arise from the constitutive properties of matter alone, without modifying the gravitational action and without introducing a new asymptotic charge. The mechanism is not tied to this particular metric: an integral equation of state combined with a geometrical condition is a constructive route to exact solutions, and it extends naturally to other closures.

The central result is the identification of finite-range hair as a distinct class of strong-field structure. It is invisible to the Gauss law at infinity and reduces to a shift of the mass when the range becomes unbounded. A single principle organizes its consequences. What matters is not the bare amplitude of the Yukawa term but the fraction of it that survives at the scale of the photon sphere, so amplitude and range enter through one effective combination. This principle accounts for the horizon structure and its dependence on both parameters, for the viability of the matter content on the negative-amplitude branch, and for the response of the horizon thermodynamics to variations of the screened sector. It also explains why the shadow and the ringdown, which are controlled by the same unstable null orbit, move together and return to RN under strong screening. The thermodynamic, optical and spectral sectors therefore do not behave as independent observables but as projections of one screening scale.

This principle has a direct observational counterpart. Confronted with the Event Horizon Telescope measurements of M87* and Sgr A*, the shadow excludes the long-range, large-amplitude corner of the parameter space and leaves a broad, strongly screened region compatible with both sources. The quasinormal spectra of scalar and electromagnetic test fields, computed with the sixth-order WKB expansion and confirmed at the sub-percent level by an independent hyperboloidal pseudospectral calculation that includes the overtones, show that the same configurations leave a correlated imprint on the ringdown. The scalar and electromagnetic sectors respond differently at finite multipole number and merge in the eikonal limit, which provides a spin-dependent handle on the shape of the potential in addition to the geometric one. Because the optical and spectral signatures originate in the same strong-field region, their joint use offers a route to distinguish screened hair from a mere rescaling of the mass and charge, which a single observable cannot do.

Several directions follow. The ringdown analysis presented here is carried out for test fields on the fixed background. Supplying a covariant action for the nonlocal sector would allow its perturbations to be coupled to those of the metric and the Maxwell field and would extend the study to a full gravitational stability analysis, in which the eikonal regime has proved decisive in other settings~\cite{Konoplya:2008ix,Konoplya:2017lhs,Konoplya:2008au}. As gravitational-wave ringdown measurements improve, the correlation between shadow and ringdown identified here can be promoted to a joint constraint on the finite-range sector, in the spirit of the correspondence between quasinormal modes and grey-body factors~\cite{Konoplya:2024lir}. 
}

\appendix
{
\section{Hyperboloidal pseudospectral computation of the quasinormal spectrum}\label{app:spectral}

This appendix gives the formulation used for the independent spectral calculation of Sec.~\ref{sec:qnm}, whose results are compared with the sixth-order WKB values in Tables~\ref{tab:qnm_scalar_main} and~\ref{tab:qnm_overtones}.

\subsection{Hyperboloidal master equation}

For the scalar ($s=0$) and electromagnetic ($s=1$) test fields, Eq.~\eqref{masterwave} can be written in the Schwarzschild-like time $t$ as
\begin{equation}
\frac{1}{f}\,\partial_t^2\Psi=\partial_r\!\left(f\,\partial_r\Psi\right)-U_s\,\Psi,
\qquad
U_s=\frac{V_s}{f}=\frac{\ell(\ell+1)}{r^2}+\left(1-s^2\right)\frac{f'}{r}.
\label{app:master}
\end{equation}
Let $r_+$ be the event horizon and $\kappa=f'(r_+)/2$ its surface gravity. We replace $t$ by the hyperboloidal time $\tau=t-h(r)$, with
\begin{equation}
h'(r)=\frac{1-g(r)}{f(r)},
\qquad
g(r)=\frac{r_+}{\kappa}\,\frac{f(r)}{r\,(r-r_+)} .
\label{app:slicing}
\end{equation}
Since $g(r_+)=2$ and $g\to0$ as $r\to\infty$ (as $r^{-2}$), one has $h\to-r_*$ at the horizon and $h\to r_*$ at infinity. The slices $\tau=\text{const}$ therefore cross the future horizon and reach future null infinity, instead of spatial infinity, and no tortoise coordinate is needed. Setting $\Psi(\tau,r)=e^{-i\omega\tau}\psi(r)$ in Eq.~\eqref{app:master} gives the quadratic eigenvalue problem
\begin{equation}
\left(L_0+\omega L_1+\omega^2L_2\right)\psi=0,
\label{app:quad}
\end{equation}
with
\begin{align}
L_0\psi&=\left(f\psi'\right)'-U_s\,\psi,\nonumber\\
L_1\psi&=i\left[2\left(1-g\right)\psi'-g'\psi\right],
\label{app:ops}\\
L_2&=\frac{g\left(2-g\right)}{f}.\nonumber
\end{align}
The function $L_2$ is regular at the horizon, where $g(2-g)$ vanishes linearly together with $f$. The quasinormal boundary conditions are not imposed explicitly. At both ends the principal part $f\,\psi''$ degenerates, and the quasinormal modes are the eigenfunctions that are regular at $r=r_+$ and at future null infinity.

\subsection{Discretization}

We use the compact coordinate $x=r_+/r\in(0,1]$, so that $\partial_r=-(x^2/r_+)\,\partial_x$. The eigenfunction is sampled at the $N$ Chebyshev nodes of the first kind mapped to $(0,1)$,
\begin{equation}
x_j=\frac12\left[1+\cos\frac{(2j-1)\pi}{2N}\right],\qquad j=1,\dots,N,
\end{equation}
which do not include the endpoints. The differentiation matrix follows from barycentric interpolation~\cite{Berrut:2004},
\begin{equation}
D_{ij}=\frac{w_j/w_i}{x_i-x_j}\ (i\neq j),\qquad
D_{ii}=-\sum_{j\neq i}D_{ij},\qquad
w_j=(-1)^j\sin\frac{(2j-1)\pi}{2N},
\end{equation}
and $\partial_r^2$ is obtained as the square of the matrix representing $\partial_r$. Equation~\eqref{app:quad} then becomes a $N\times N$ matrix polynomial, which we linearize in the usual way by introducing $v=(\psi,\omega\psi)^{\rm T}$,
\begin{equation}
\begin{pmatrix}0&\mathbb 1\\-L_0&-L_1\end{pmatrix}v=
\omega\begin{pmatrix}\mathbb 1&0\\0&L_2\end{pmatrix}v ,
\end{equation}
and solve as a dense generalized eigenproblem. Eigenvalues with $\mathrm{Re}\,\omega>0$ and $\mathrm{Im}\,\omega<0$ are retained. Genuine modes are identified by their stability under changes of $N$: the physical eigenvalues do not move, whereas the spurious ones associated with the finite discretization do.

\subsection{Tests}

As a check of the implementation, for $Q=l=0$ and $M=1$ we obtain, with $N=60$, $M\omega=0.483644-0.096759\,i$ ($n=0$) and $0.463851-0.295604\,i$ ($n=1$) for the $\ell=2$ scalar field, which coincide with the Schwarzschild values to the six digits shown. For the Yukawa geometries we computed each frequency with $N=50,60,70$. Table~\ref{tab:spectral_conv} lists the largest relative variation $\max_N|\omega_N-\omega_{60}|/|\omega_{60}|$ for the configurations of Table~\ref{tab:qnm_overtones}. The fundamental mode is stable to $10^{-9}$ and the first overtone to $10^{-6}$; the second overtone, which is the most sensitive to the discretization, is stable to a few parts in $10^{3}$ at most.

\begin{table}[h]
\centering
\footnotesize\setlength{\tabcolsep}{3pt}
\caption{{Relative variation of the spectral frequencies under $N=50,60,70$ for $\ell=3$, $M=1$ and $Q=0.5$.}}
\label{tab:spectral_conv}
\begin{ruledtabular}
\begin{tabular}{ccccccc}
 & \multicolumn{3}{c}{Scalar} & \multicolumn{3}{c}{Electromagnetic}\\
Geometry & $n=0$ & $n=1$ & $n=2$ & $n=0$ & $n=1$ & $n=2$\\
\hline
RN & $1.4\times10^{-9}$ & $1.1\times10^{-6}$ & $4.0\times10^{-3}$ & $1.1\times10^{-9}$ & $4.2\times10^{-7}$ & $9.3\times10^{-4}$\\
$A=1,\ l=-1$ & $1.5\times10^{-9}$ & $2.5\times10^{-7}$ & $1.6\times10^{-4}$ & $1.5\times10^{-9}$ & $1.7\times10^{-7}$ & $1.4\times10^{-4}$\\
$A=4,\ l=-1$ & $1.4\times10^{-9}$ & $3.6\times10^{-7}$ & $9.8\times10^{-4}$ & $8.9\times10^{-10}$ & $8.4\times10^{-7}$ & $1.1\times10^{-3}$\\
\end{tabular}
\end{ruledtabular}
\end{table}
}

\section*{ACKNOWLEDGMENTS}

Y. Gómez-Leyton acknowledges ANID, Subvención a la Academia, convocatoria año 2025, Grant No.~85250186.

\bibliography{biblio.bib}
\bibliographystyle{elsarticle-num}

\end{document}